\documentclass[aps,prd,preprintnumbers,twocolumn,groupedaddress,nofootinbib,showpacs]{revtex4}
\usepackage{graphicx}
\def\beq{\begin{equation}}
\def\eeq{\end{equation}}
\def\bey{\begin{eqnarray}}
\def\eey{\end{eqnarray}}

\def\lsim{\mathrel{\raise.3ex\hbox{$<$\kern-.75em\lower1ex\hbox{$\sim$}}}}
\def\gsim{\mathrel{\raise.3ex\hbox{$>$\kern-.75em\lower1ex\hbox{$\sim$}}}}

\begin{document}

\title{Pulsars Cannot Account for the Inner Galaxy's GeV Excess}  
\author{Dan Hooper$^{1,2}$}
\author{Ilias Cholis$^{1}$}
\author{Tim Linden$^3$}
\author{Jennifer Siegal-Gaskins$^4$}
\author{Tracy Slatyer$^5$}
\affiliation{$^1$Center for Particle Astrophysics, Fermi National Accelerator Laboratory, Batavia, IL 60510}
\affiliation{$^2$Department of Astronomy and Astrophysics, University of Chicago, 5640 S Ellis Ave., Chicago, IL 60637}
\affiliation{$^3$Department of Physics, University of California, Santa Cruz, 1156 High Street, Santa Cruz, CA 95064}
\affiliation{$^4$Einstein Fellow, California Institute of Technology, 1200 E.~California Blvd., Pasadena, CA 91125}
\affiliation{$^5$School of Natural Sciences, Institute for Advanced Study, Princeton, NJ 08540}

\date{\today}

\begin{abstract}

Using data from the Fermi Gamma-Ray Space Telescope, a spatially extended component of gamma rays has been identified from the direction of the Galactic Center, peaking at energies of $\sim$2-3 GeV. More recently, it has been shown that this signal is not confined to the innermost hundreds of parsecs of the Galaxy, but instead extends to at least $\sim$3 kpc from the Galactic Center. While the spectrum, intensity, and angular distribution of this signal is in good agreement with predictions from annihilating dark matter, it has also been suggested that a population of unresolved millisecond pulsars could be responsible for this excess GeV emission from the Inner Galaxy. In this paper, we consider this later possibility in detail. Comparing the observed spectral shape of the Inner Galaxy's GeV excess to the spectrum measured from 37 millisecond pulsars by Fermi, we find that these sources exhibit a spectral shape that is much too soft at sub-GeV energies to accommodate this signal. We also construct population models to describe the spatial distribution and luminosity function of the Milky Way's millisecond pulsars. After taking into account constraints from the observed distribution of Fermi sources (including both sources known to be millisecond pulsars, and unidentified sources which could be pulsars), we find that millisecond pulsars can account for no more than $\sim$10\% of the Inner Galaxy's GeV excess. Each of these arguments strongly disfavor millisecond pulsars as the source of this signal.

\end{abstract}

\pacs{97.60.Gb, 95.55.Ka, 95.35.+d; FERMILAB-PUB-13-129-A}
\maketitle




\section{Introduction}

Over the past few years, a number of groups using data from the Fermi Gamma-Ray Space Telescope have identified a spatially extended component of gamma rays in the region of the Galactic Center, peaking at energies of $\sim$2-3 GeV~\cite{Hooper:2011ti,Abazajian:2012pn,HG2,HG1,Boyarsky:2010dr}. This signal has been interpreted as possible evidence of annihilating dark matter particles, and can be well fit by 7-12 GeV particles annihilating to $\tau^+\tau^-$ (possibly among other leptons)~\cite{Hooper:2011ti,Abazajian:2012pn,HG2} or by 22-45 GeV particles annihilating to quarks~\cite{Hooper:2011ti,Abazajian:2012pn,HG2,HG1} (alternatively, see Ref.~\cite{Hooper:2012cw}). In either case, the morphology of the gamma-ray signal requires a dark matter distribution which scales approximately as $\rho \propto r^{-1.2}\,\,\, {\rm to}\,\,\, r^{-1.4}$, where $r$ is the distance to the Galactic Center~\cite{Hooper:2011ti,Abazajian:2012pn}, and an annihilation cross section that is on the order of a few times $\sigma v\sim 2 \times 10^{-27}$ cm$^3$/s~\cite{Hooper:2011ti} (up to overall uncertainties in the normalization of the halo profile). This required dark matter distribution is  in good agreement with expectations based on hydrodynamical simulations (see Ref.~\cite{Gnedin:2011uj} and references therein) and is consistent with current observational constraints~\cite{Iocco:2011jz}. Similarly, the required annihilation cross section is comfortably within the range expected for a thermal relic of the Big Bang. 

Until recently, most opinions regarding the Galactic Center's GeV gamma-ray signal fell within one of three broad categories. First were those who found the dark matter interpretation to be compelling, and considered the gamma-ray signal to be difficult to explain with astrophysical sources or backgrounds. Second were those who argued that a population of a few thousand millisecond pulsars concentrated very densely within the inner tens of parsecs of the Galaxy were at least as likely to be responsible for the observed gamma-ray emission (as discussed in Refs.~\cite{HG2,Hooper:2011ti,Wharton:2011dv,Abazajian:2012pn} and advocated for in Ref.~\cite{pulsars}). And third, is the not uncommon view that the region of the Galactic Center is too astrophysically complex to reliably isolate or identify any gamma-ray signal that might result from annihilating dark matter particles. 

Very recently, however, this situation has changed substantially. In Ref.~\cite{tracy}, an analysis of Fermi data was presented which identified a component of gamma rays from the inner kiloparsecs of the Milky Way, exhibiting a spectrum and morphology consistent with the signal previously observed from the Galactic Center. But whereas earlier studies were able to confidently identify this signal only within a few hundred parsecs of the Galactic Center, the template technique applied in Ref.~\cite{tracy} was able to clearly detect this signal out to at least $\sim$2-3 kpc to the north and south, and with evidence of extension out to $\sim$5 kpc~\cite{tracy}. This result is important in that it does not rely on observations of the complex region of the Galactic Center, and is not highly sensitive to uncertainties in Fermi's low energy point spread function, thus circumventing the concerns of the third group mentioned in the previous paragraph.

\begin{figure*}
\mbox{\includegraphics[width=0.7\textwidth,clip]{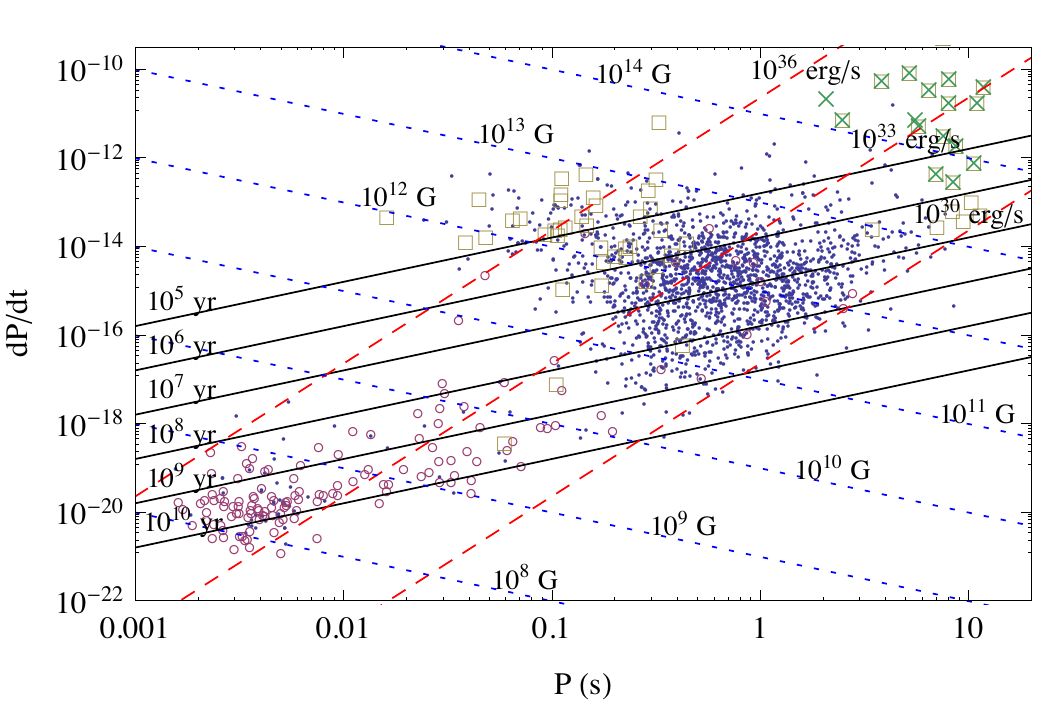}}
\caption{The period ($P$) and its rate of change ($dP/dt$), for pulsars described in the ATNF catalog~\cite{atnf}. Open circles denote binary pulsars (most of which are millisecond pulsars), while squares and $X$'s represent radio quiet pulsars and anomalous X-ray and/or soft gamma-ray emitters, respectively. All other pulsars are shown as dots. Also shown are contours of constant spin-down power (red dashes), magnetic field (blue dots), and characteristic age, $\tau\equiv P/2\dot{P}$ (solid black), calculated assuming a neutron star mass of $1.4 M_{\odot}$ and a radius of 10 km. See text for discussion.}
\label{PPdot}
\end{figure*}

In this paper, we discuss whether this GeV excess observed from the Inner Galaxy~\cite{tracy} could potentially be explained by a population of gamma-ray pulsars. Here, and throughout this paper, we use the phrase ``Inner Galaxy'' to refer to the region within several kiloparsecs of the Galactic Center, as opposed to the much smaller region ($\sim$100 pc) referred to as the ``Galactic Center''. After briefly reviewing some of the most relevant aspects of pulsars in Sec.~\ref{review}, we turn our attention in Sec.~\ref{spec} to the gamma-ray spectra observed from known millisecond pulsars. We find that the spectrum observed from these sources is not compatible with that of the Inner Galaxy's GeV excess. Independently of spectral arguments, we demonstrate in Sec.~\ref{dist} that the Inner Galaxy's GeV excess cannot be accounted for with millisecond pulsars without significantly overpredicting the number of pulsars that Fermi would have resolved as individual point sources. Pulsar distributions which are consistent with Fermi's source catalog can account for no more than $\sim$10\% of the observed GeV excess.  From either of these arguments, we conclude that gamma-ray pulsars cannot account for a significant fraction of the Inner Galaxy's GeV excess. In Sec.~\ref{summary}, we briefly summarize our results and their implications.

\section{A Brief Review of Ordinary and Millisecond Pulsars}
\label{review}

Pulsars are rapidly spinning neutron stars which steadily convert their rotational kinetic energy into radiation, including potentially observable emission at radio and gamma-ray wavelengths. When initially formed, pulsars typically exhibit rotational periods on the order of tens or hundreds of milliseconds, and magnetic field strengths of $\sim$$10^{11}$-$10^{13}$ G (see Fig.~\ref{PPdot}). As a result of magnetic-dipole braking, a pulsar's period will slow down at a rate given by $\dot{P} = 3.3\times 10^{-15} \,(B/10^{12}\, {\rm G})^2 \, (P/0.3 \,{\rm s})^{-1}$, corresponding to an energy loss rate of $\dot{E}=4 \pi^2 I \dot{P}/P^3 = 4.8\times 10^{33} {\rm erg/s} \, (B/10^{12}\, {\rm G})^2 \, (P/0.3 \,{\rm s})^{-4} \, (I/10^{45} \rm{g}\,\rm{cm}^2)$. As a result of this rotational slowing, pulsars steadily become less luminous. For very young pulsars, this occurs very rapidly. Within a few hundred thousand years, the Crab and Vela pulsars will become a thousand times fainter than they are at present. After $\sim$$10$-$100$ million years, such objects slow to a point at which they are no longer able to generate radio emission (crossing what is known as the pulsar ``death line'').

To those pulsars that are gravitationally bound to a binary companion, another stage of evolution is possible. If at some point in time (likely well after the pulsar has lost most of its rotational kinetic energy and become faint) the companion enters a red giant phase, accretion onto the pulsar and the corresponding transfer of angular momentum can dramatically increase the rotational speed of the pulsar (to $P$$\sim$1.5-100 ms), while also dramatically reducing the magnetic field (to $B$$\sim$$10^8$-$10^9$ G). Such millisecond pulsars (MSPs) (also known as spun-up or recycled pulsars) can be as luminous as ordinary pulsars, but evolve much more slowly, remaining bright for billions of years. 

Another important difference between ordinary pulsars and MSPs is found in their velocity distributions.  In the process of their formation, pulsars receive substantial kick velocities, resulting from small asymmetries in their collapse. The average velocity observed among young pulsars is $\sim$$400\pm40$ km/s~\cite{hobbs}, which is much higher than is observed among other stellar populations. MSPs, however, necessarily consist of neutron stars that have either retained or captured a binary companion, and thus must have had unusually weak kick velocities. Furthermore, the additional mass of the companion star further reduces the velocities acquired by these systems. In Appendix~\ref{kicks}, we discuss this in more detail and estimate that instead of average kick velocities of $\sim$400 km/s, MSPs should receive average velocities of 10-50 km/s, with the precise value depending on the details of the stellar distribution being considered. This conclusion is supported by the much lower velocities that are observed among MSPs~\cite{lyne}, and by the fact that most observed MSPs reside within globular clusters, which have escape velocities on the order of only tens of km/s.

MSPs are of particular interest in this study for two reasons. First, the morphology of the gamma-ray signal from the Galactic Center is highly concentrated, much more so than is expected from the overall stellar distribution. In particular, if any stellar population is to produce this signal, its members must preferentially be located very centrally around the inner tens of parsecs surrounding the Galactic Center, with a number density that scales approximately as $n(r) \propto r^{-2.4}$ to $r^{-2.8}$~\cite{Hooper:2011ti,Abazajian:2012pn}.\footnote{Note that for dark matter, the annihilation rate per volume is proportional to the {\it square} of the dark matter density, and thus the observed morphology of the Galactic Center's gamma-ray signal favors $\rho_{\rm DM} \propto r^{-1.3}$, rather than the $n \propto r^{-2.6}$ required of a stellar distribution.} This extremely steep distribution would be very difficult to accommodate with ordinary pulsars, whose kick velocities are more than sufficient to expel the overwhelming majority of pulsars from the Galactic Center. With their much weaker kick velocities, however, MSPs which form in the Galactic Center could potentially remain concentrated in this region. Furthermore, the large numbers of MSPs present in globular clusters suggest that they are produced in part as a result of stellar encounters (see Sec.~3.3 of Ref.~\cite{glob}, and reference therein).  If this conclusion also applies to the Galactic Center, it could explain why the morphology of the observed gamma-ray signal is so much more centrally concentrated than the overall stellar distribution. Second, the gamma-ray emission identified in Ref.~\cite{tracy} extends to at least $\sim$3 kpc north and south of the Galactic Center. By the time that a newly formed ordinary pulsar could travel more than a few hundred parsecs from the location of its birth (likely near the Galactic Plane), it will have lost the vast majority of its initial rotational energy, and become too faint to significantly contribute to the observed gamma-ray emission. Observations also support the fact that ordinary pulsars are found overwhelmingly within or near the volume of the stellar disk. In contrast, MSPs can remain bright for billions of years. Thus if a significant number of MSPs somehow escape the gravitational potential of their environment, there could potentially exist a population of luminous, high-latitude gamma-ray pulsars.

\section{Comparison with the observed Gamma-Ray Spectra of Millisecond Pulsars}
\label{spec}

\begin{figure}
\mbox{\includegraphics[width=0.49\textwidth,clip]{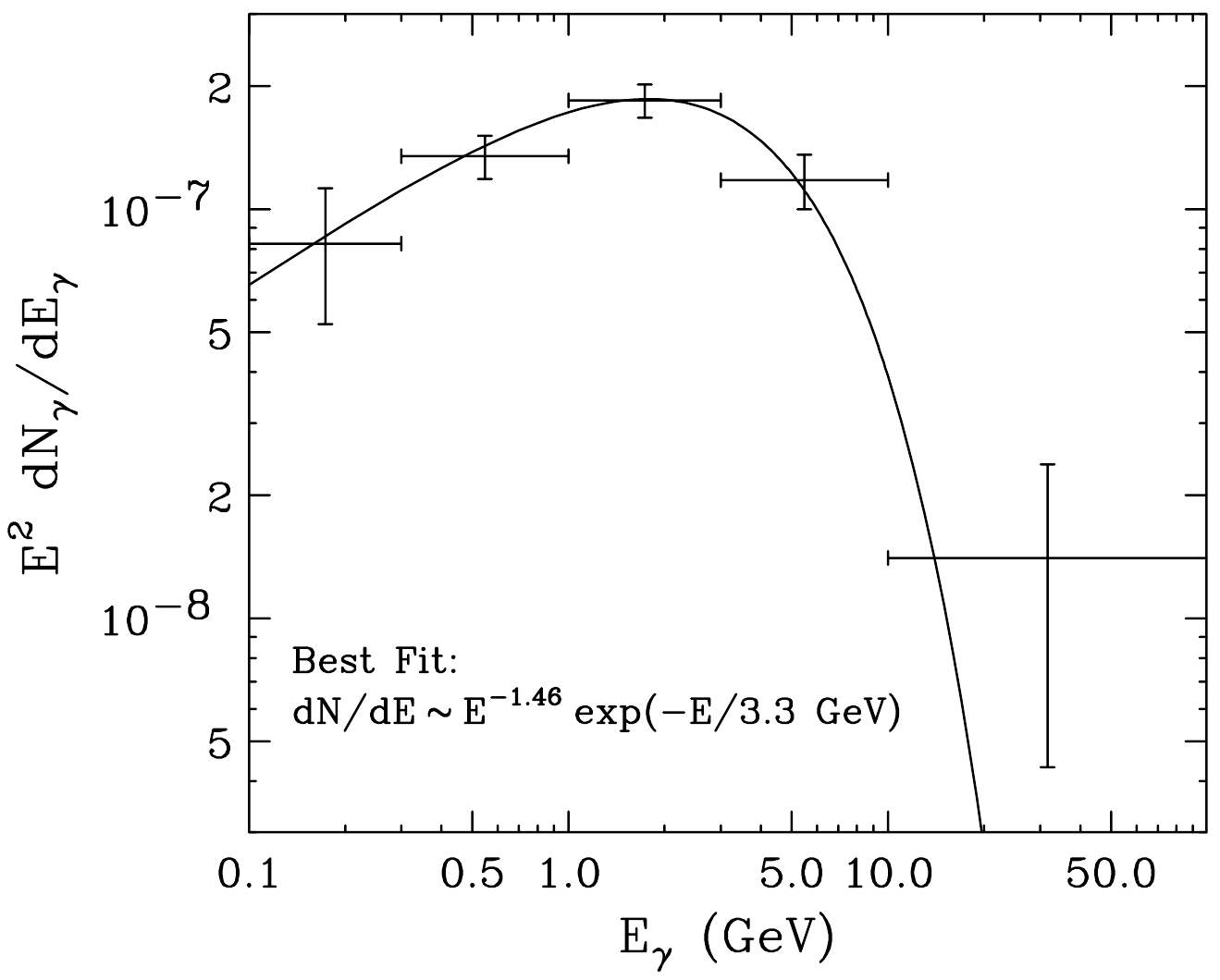}}
\caption{The combined gamma-ray spectrum from 37 millisecond pulsars observed by Fermi. The solid line shows the best-fit parametrization to this spectrum, $dN_{\gamma}/dE_{\gamma} \propto E_{\gamma}^{-1.46} \exp(-E_{\gamma}/3.3\,{\rm GeV})$.}
\label{avespec}
\end{figure}

The Fermi Collaboration has detected gamma-ray emission from a total of 125 sources identified as pulsars, 47 of which have millisecond-scale periods~\cite{publiclist}. Of these 47 MSPs, 37 have spectral information listed in the second Fermi source catalog (2FGL)~\cite{catalog}. In Fig.~\ref{avespec}, we plot the sum of the spectra of these 37 sources. We find that this collection of sources can be very well fit by the standard pulsar spectral parametrization, $dN_{\gamma}/dE_{\gamma} \propto E_{\gamma}^{-\Gamma} \exp(-E_{\gamma}/E_{\rm cut})$, with best-fit values of $\Gamma=1.46$ and $E_{\rm cut}=3.3$ GeV. This is very similar to the results found in earlier studies, based on a smaller number of Fermi MSPs~\cite{averagepulsar,Malyshev:2010xc}.

\begin{figure}
\mbox{\includegraphics[width=0.49\textwidth,clip]{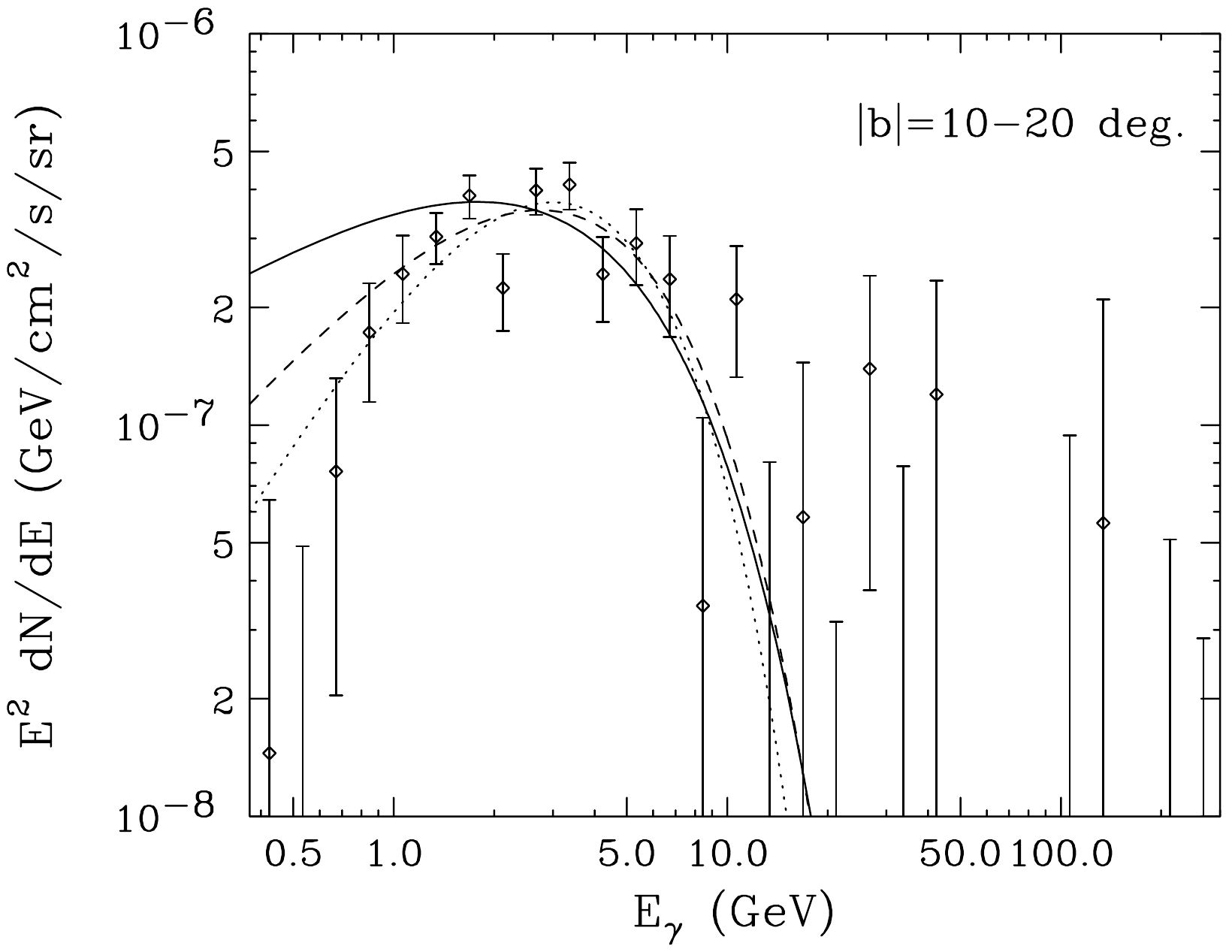}}
\caption{The gamma-ray spectrum of the $|b|=10^{\circ}-20^{\circ}$ regions of the Inner Galaxy (see Ref.~\cite{tracy}), after subtracting emission from inverse Compton scattering~\cite{tracy}, compared to the spectral shape best-fit to 37 MSPs observed by Fermi (solid line); see Fig.~\ref{avespec}. Also shown for comparison are the shapes corresponding to spectral parameters that better match this emission: $\Gamma=1.0$, $E_{\rm cut}=2.75$ GeV (dashed) and $\Gamma=0.5$, $E_{\rm cut}=2.0$ GeV (dotted).}
\label{bubblespec}
\end{figure}


In Fig.~\ref{bubblespec}, we compare this best-fit spectrum (solid line) to that observed from the $|b|=10^{\circ}-20^{\circ}$ regions of the Inner Galaxy, after subtracting an inverse Compton component that accounts for the Fermi Bubbles emission (see Ref.~\cite{tracy} for details).\footnote{Note that while we have chosen to compare to the spectrum observed from the $|b|=10^{\circ}-20^{\circ}$ region of the Inner Galaxy, this spectrum is very similar to that extracted from higher latitude regions~\cite{tracy}. We have chosen to not use the spectrum extracted from the $|b|=1^{\circ}-10^{\circ}$ region due to difficulties in separating this signal from emission associated with the Galactic Disk.} The spectral shape observed from these 37 resolved MSPs exhibits a much softer spectral index than the spectrum of the excess emission observed from the Inner Galaxy, especially at energies below $\sim$1-2 GeV. Also shown for comparison are harder spectral shapes, corresponding to $\Gamma=1.0$, $E_{\rm cut}=2.75$ GeV (dashes) and $\Gamma=0.5$, $E_{\rm cut}=2.0$ GeV (dots). While such hard spectra can provide a good fit to the emission observed from the Inner Galaxy (especially the $\Gamma=0.5$, $E_{\rm cut}=2.0$ GeV case), they are not consistent with the spectral shape shown in Fig.~\ref{avespec}. The comparison between these harder spectral shapes and the error bars shown in Fig.~\ref{avespec} yields fits of $\chi^2=17.8$ and 38.9 (over 5-1 degrees-of-freedom) for these two parameter sets ($\Gamma=0.5$, $E_{\rm cut}=2$ GeV, and $\Gamma=1.0$, $E_{\rm cut}=2.75$ GeV, respectively), each of which can be excluded at beyond the 99.8\% confidence level. At least at face value, it appears that we can exclude at high confidence a MSP origin for the emission observed from the low-latitude regions of the Inner Galaxy.

\begin{figure*}
\mbox{\includegraphics[width=0.7\textwidth,clip]{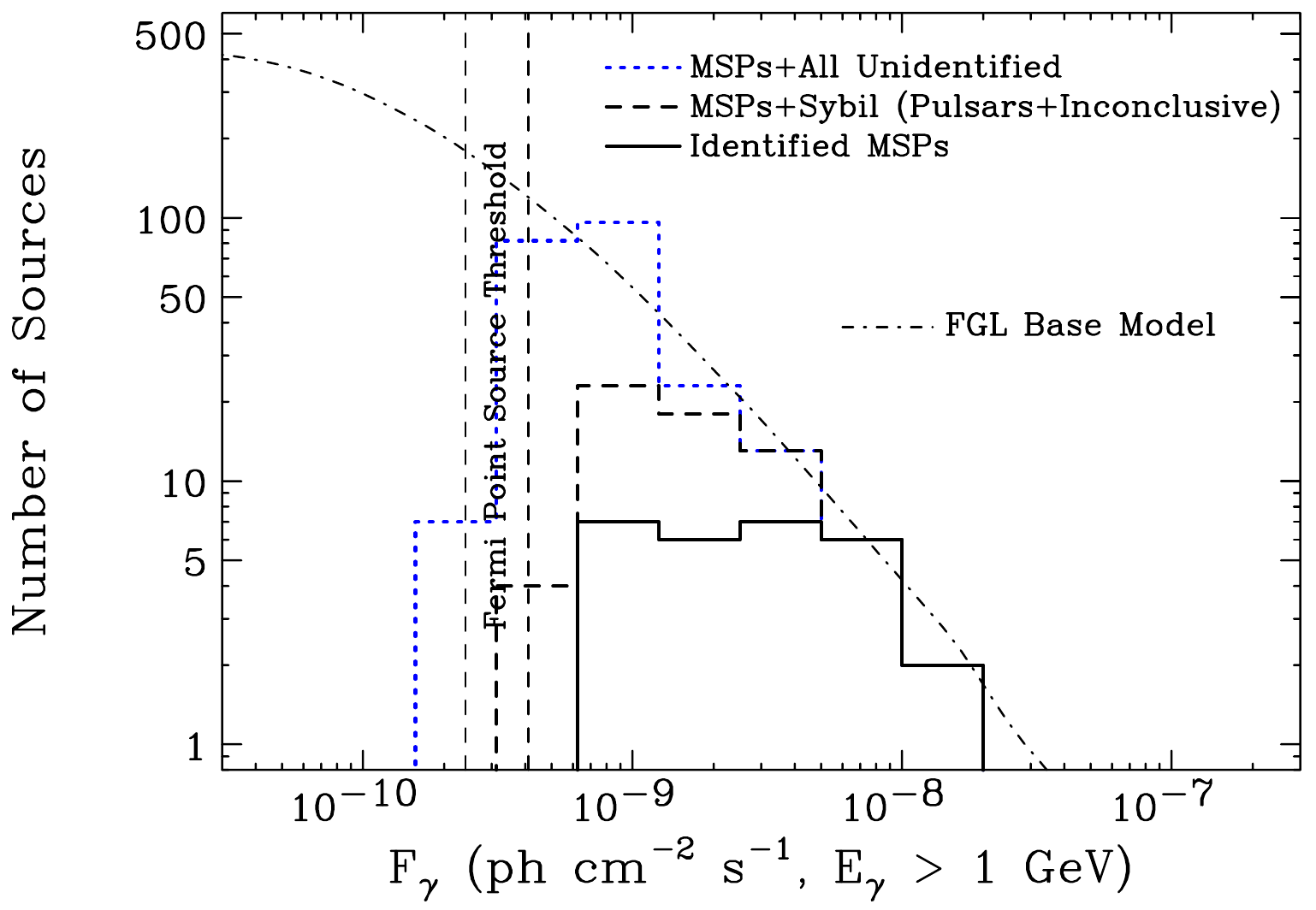}}
\caption{The observed flux distribution (proportional to $dN/d\log S$) of identified millisecond pulsars with $|b|>10^{\circ}$ (solid black), compared to that predicted in the base model of Ref.~\cite{loeb} ($\langle|z|\rangle = 1$ kpc, $\sigma_r=5$ kpc, $B_0=10^{8}$ G, normalized to accomodate the observed number of very bright sources). Also shown are the distributions of identified MSPs plus all unidentified Fermi sources (dotted blue), and of identified MSPs plus all unidentified sources found by the {\it Sibyl} algorithm~\cite{sibyl} to be either likely pulsars or sources of an inconclusive nature (dashed black). Also shown is the range of Fermi's threshold to resolve an individual source~\cite{catalog}. This base model cannot account for the observed number of bright MSPs without significantly overpredicting the number of fainter MSPs.}
\label{histogram-dist}
\end{figure*}

Perhaps, however, the MSPs that have been resolved by Fermi are not representative of all such objects, and the diffuse emission from the sum of all unresolved (and faint) MSPs has a much harder spectral index than is observed from resolved (and brighter) sources. Among Fermi's MSPs, however, we see no evidence for this. In particular, we find a best-fit spectral index of $\Gamma=1.36$ for the sum of the 21 MSPs with gamma-ray luminosities less than $10^{37}$ GeV/s (above 100 MeV), and $\Gamma=1.34$ for the sum of the 8 MSPs with gamma-ray luminosities less than $10^{36}$ GeV/s. These values are not very different from that found in our overall best-fit, $\Gamma=1.46$. If MSPs exhibit a correlation between hard spectral indices and low luminosities, this trend is not evident among the observed source population. 

Furthermore, if hard spectrum, low-luminosity sources dominate the diffuse emission from MSPs, then the hard spectral index should be reflected in the emission observed from globular clusters, which should contain a representative sample of MSPs. Although the spectra of the 11 globular clusters included in the 2FGL~\cite{catalog} have large error bars and are thus difficult to evaluate individually, the sum of the spectra from these 11 sources is quite similar to that observed from Fermi's MSPs. Similarly, the Fermi Collaboration studied 8 globular clusters and found their (statistically weighted) average spectral index to be $\Gamma=$1.35~\cite{glob}, again similar to that observed from resolved MSPs. 


Although we have shown in this section that the gamma-ray spectrum observed from individual MSPs (and from globular clusters) is incompatible with that from the Inner Galaxy as reported in Ref.~\cite{tracy}, one might worry that systematic uncertainties in the low-energy ($\lsim$\,1 GeV) spectrum could possibly alter this conclusion. The error bars presented in Ref.~\cite{tracy} (and shown in our Fig.~\ref{bubblespec}) are purely statistical, and do not reflect the possible mismodeling of point sources or of diffuse emission. While the over-subtraction of low-energy emission from known point sources could, in principle, lead to an artifically hard spectrum at low-energies, only if the Fermi collaboration's source catalog~\cite{Abdo:2010ru} overestimates the total flux from the 35 sources in the $|b|=10^{\circ}-20^{\circ}$ region, for example, by more than a factor of two in the in the 300-1000 MeV range (a variation several times larger than the quoted errors) could the spectrum of the Inner Galaxy's GeV excess be consistent with that observed from individual MSPs. More difficult to rule out is the possibily that the Fermi collaboration's diffuse model significantly overestimates the density of cosmic rays in the region of interest, leading the analysis of Ref.~\cite{tracy} to effectively oversubtract gamma-ray emission from pion production and other diffuse processes, potentially artificially hardening the spectrum of the GeV excess at low energies. We note that the fit residuals from the analysis of Ref.~\cite{tracy}, averaged over the regions in question, are much smaller than the signal at all relevant energies; re-adding them to the signal does not meaningfully soften the spectrum.

Although systematic uncertainties in the Fermi instrument response functions below 1 GeV could plausibly skew the inferred spectral shape, no evidence for this is seen in other spectral components, such as that associated with the Fermi diffuse model~\cite{tracy}; this argues against an energy-dependent error in Fermi's effective area being responsible for the apparently hard spectrum of the Inner Galaxy's GeV excess. Furthermore, the large angular size of the regions of interest, and their significant distance from the Galactic Center, make any mismodeling of Fermi's point spread function an unlikely source of large distortions to the spectrum.


To summarize the results of this section, we find that the gamma-ray spectra observed from individual MSPs consistently reveal a spectral index that is much too soft to accommodate the signal observed from the Inner Galaxy. Furthermore, we find no evidence for a population of low-luminosity and spectrally hard MSPs that might be able to account for the signal.

\section{The Distribution of Millisecond Pulsars in the Milky Way}
\label{dist}

In the previous section, we showed that the gamma-ray spectrum observed from individual MSPs (and from collections of MSPs in globular clusters) is not consistent with the spectral shape of the Inner Galaxy's GeV excess. In this section, we set aside this conclusion for the time being and focus instead on constraints derived from the observed spatial and flux distributions of MSPs. We will use this information to assess the question of whether the intensity and morphology of the Inner Galaxy's GeV excess might originate from a population of unresolved MSPs.

\subsection{Millisecond Pulsars Associated with the Galactic Disk}

We begin by considering MSPs which follow a distribution similar to that of the Milky Way's disk. As our starting point, we adopt the ``base model'' of Ref.~\cite{loeb}, which includes a spatial distribution and luminosity function for MSPs in the Milky Way. In particular, we adopt a spatial distribution of MSPs with a number density given by:
\begin{equation}
n(r,z) \propto \exp(-r^2/2\sigma^2_r) \, \exp(-|z|/\langle |z| \rangle),
\label{spatial}
\end{equation}
where $r$ and $z$ describe the location in cylindrical coordinates. To begin, we will consider values of $\sigma_r=5$ kpc and $\langle |z| \rangle=1$ kpc, as adopted in the ``base model'' of Ref.~\cite{loeb}.


Again following Ref.~\cite{loeb}, we take the gamma-ray luminosity (above 100 MeV) of a MSP to be equal to 5\% of its energy loss rate, $\dot{E}$, except for the most luminous sources which follow $L_{\gamma} \propto \sqrt{\dot{E}}$. For the distribution of MSP periods, we assume $dN/dP \propto P^{-2}$, with a minimum value of 1.5 msec~\cite{period} (the most rapidly spinning pulsar observed to date has a period of 1.4 msec~\cite{rapid}). The time derivative of a MSP's period is determined by its magnetic field (through magnetic dipole braking, see Sec.~\ref{review}). The magnetic fields are taken to follow a log-normal distribution centered around $B_0=10^8$ G and with a logarithmic standard deviation of 0.2. While we take the gamma-ray spectrum to follow the form of the best-fit as shown in Fig.~\ref{avespec}, the precise spectral shape does not significantly impact any of the results presented in this section.

In Fig.~\ref{histogram-dist}, we show the flux distribution of high-latitude ($|b|>10^{\circ}$) MSPs (proportional to $dN/d\log S$) predicted in the base model of Ref.~\cite{loeb} (labeled ``FGL Base Model''). We compare this prediction to the number of sources as observed by Fermi. Here, the solid black histogram describes the distribution of sources in the 2FGL which have been identified as MSPs~\cite{publiclist}, while the dotted blue histogram denotes the sum of the identified MSPs along with all presently unidentified sources in the catalog ({\it i.e.}~all sources listed as unassociated in the 2FGL that do not appear on the list of Fermi pulsars~\cite{publiclist} and have not since been identified in Ref.~\cite{Massaro:2013uoa} as a blazar). For the predicted distribution, we have normalized the total number of MSPs to approximately match the observed number of very bright MSPs ($F_{\gamma}(>1\,\rm{GeV})\sim 10^{-8}$ ph cm$^{-2}$\,s$^{-1}$). Note that in this respect, we depart from the base model of Ref.~\cite{loeb}. For this choice of normalization, we find that unresolved MSPs in this model produce about 0.5\% of the high latitude ($|b|>40^{\circ}$) diffuse gamma-ray background at $E_{\gamma}\sim 1$ GeV, which is about a factor of three below the maximum value consistent with Fermi's anisotropy constraint~\cite{SiegalGaskins:2010nh,SiegalGaskins:2010mp}. The vertical dashed lines in Fig.~\ref{histogram-dist} denote the range of Fermi's threshold for a source out of the plane ($|b|>10^{\circ}$) to be included in the 2FGL catalog, as quoted in Ref.~\cite{catalog}. The range of this threshold spans sources with effective spectral indices between -1 and -3 (assuming a power-law form). We note that the point source threshold for MSPs should typicaly fall near the lower end of this range, since their spectra are relatively hard at $\sim$1 GeV.

\begin{figure}
\mbox{\includegraphics[width=0.49\textwidth,clip]{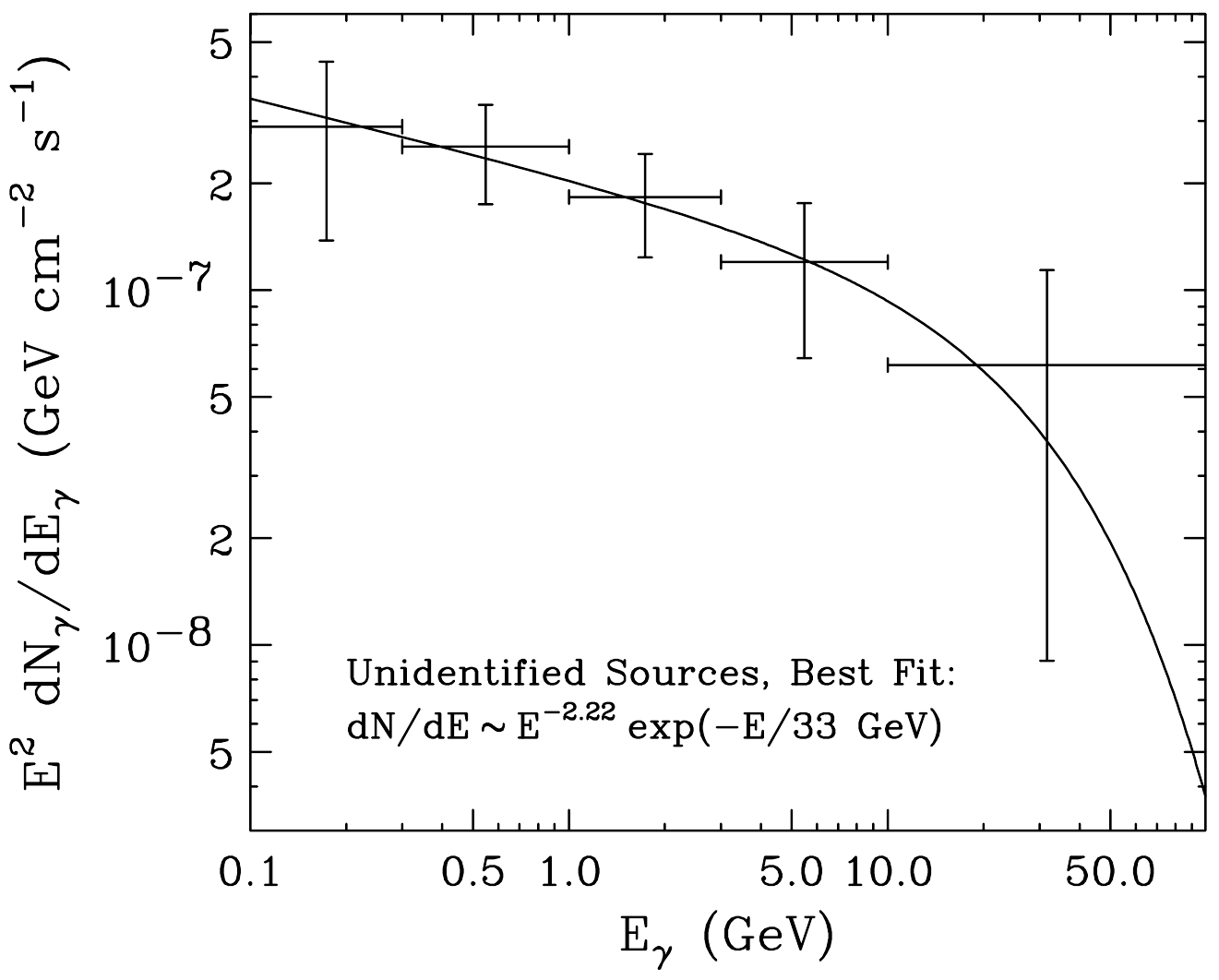}}
\caption{The combined gamma-ray spectrum from all of Fermi's unidentified sources located outside of the plane ($|b|>10^{\circ}$). The solid line shows the best-fit parametrization to this spectrum, $dN_{\gamma}/dE_{\gamma} \propto E_{\gamma}^{-2.22} \exp(-E_{\gamma}/33\,{\rm GeV})$. This spectrum does not resemble that observed from individual millisecond pulsars, but is consistent with that observed from blazars and other types of AGN.}
\label{avespecunidentified}
\end{figure}

While this model predicts a diffuse gamma-ray signal from the Inner Galaxy that is similar to the observed GeV excess (with a similar morphology, and an overall intensity that is only a factor of a few less than observed), it also significantly overpredicts the number of MSPs with $F_{\gamma}(>1\,\rm{GeV})\sim 10^{-9}$ ph cm$^{-2}$\,s$^{-1}$. Only if essentially {\it all} of Fermi's unidentified sources (above $|b|=10^{\circ}$) are MSPs could this model be potentially compatible with the observed flux distribution. It is clear, however, that only a modest fraction of these unidentified sources are pulsars, and that most of them are instead blazars or other types of active galactic nuclei (AGN). For example, the authors of Ref.~\cite{sibyl}, using the random forest classifier {\it Sibyl}, trained on the observed spectra and variability of over 900 identified Fermi point sources (AGN and pulsars), determined that at least 80\% of Fermi's unidentified high-latitude ($|b|>10^{\circ}$) sources are likely AGN. Furthermore, the overall spectrum from this collection of unidentified sources does not resemble that observed from individual MSPs (or observed from the Inner Galaxy), but instead resembles that of AGN. In Fig.~\ref{avespecunidentified}, we plot the combined spectrum of these unidentified sources (all with $|b|> 10^{\circ}$). This spectrum shows no sign of a sharp spectral peak at $\sim$2 GeV, as the Inner Galaxy's diffuse emission does, nor does it resemble the more mildly peaked spectrum observed among the identified MSPs. The shape of this spectrum strongly suggests that most of the unidentified sources are not pulsars, but are instead mostly AGN or other soft-spectrum gamma-ray sources. 

In Fig.~\ref{histogram-dist}, the black-dashed histogram represents the distribution of identified MSPs added to the distribution of sources classified by {\it Sibyl} as either a likely pulsar, or as a source of an inconclusive nature (only sources classified as likely AGN were not included in this distribution). This distribution represents an approximate upper limit for the numbers of Fermi's sources that could potentially be MSPs. In all likelihood, the true distribution falls somewhere between the solid-black histogram (presently identified MSPs) and the dashed-black histogram (identified MSPs plus {\it Sibyl}'s likely pulsars and inconclusive sources). When comparing these distributions to that predicted by the base model of Ref.~\cite{loeb}, we are forced to conclude that this model cannot account for the observed number of very bright MSPs without predicting far too many fainter MSPs.

\begin{figure*}
\mbox{\includegraphics[width=0.49\textwidth,clip]{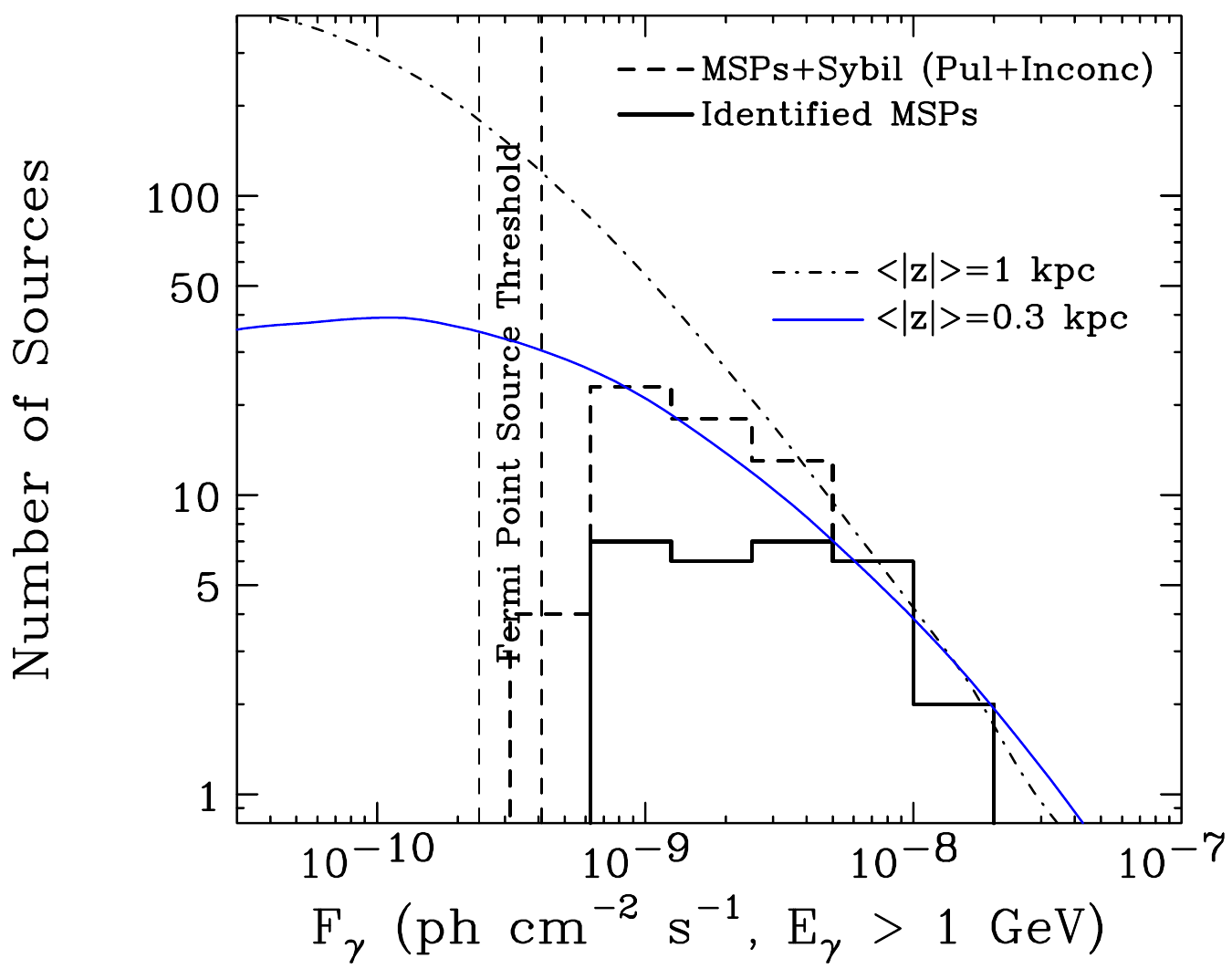}}
\mbox{\includegraphics[width=0.49\textwidth,clip]{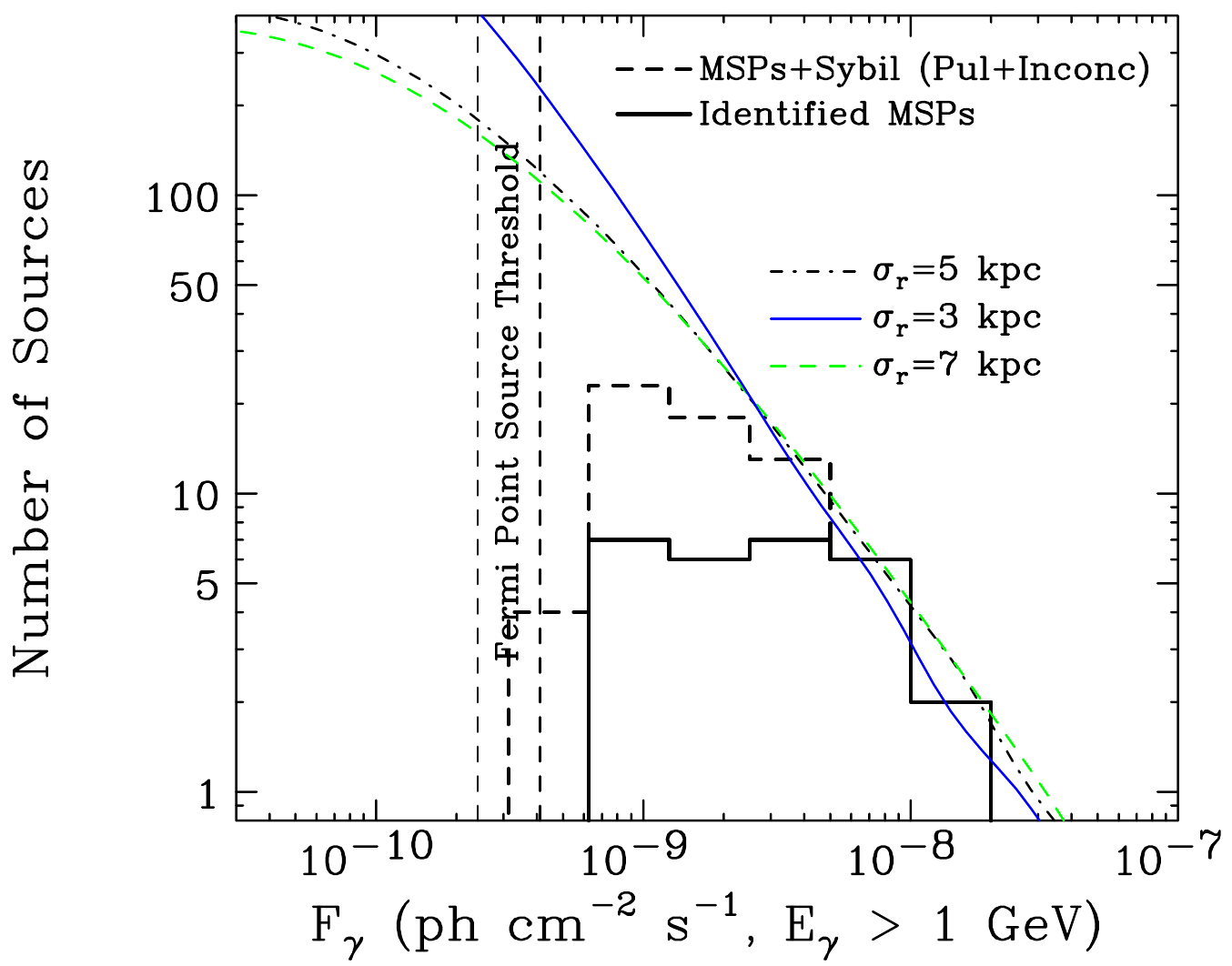}}\\
\mbox{\includegraphics[width=0.49\textwidth,clip]{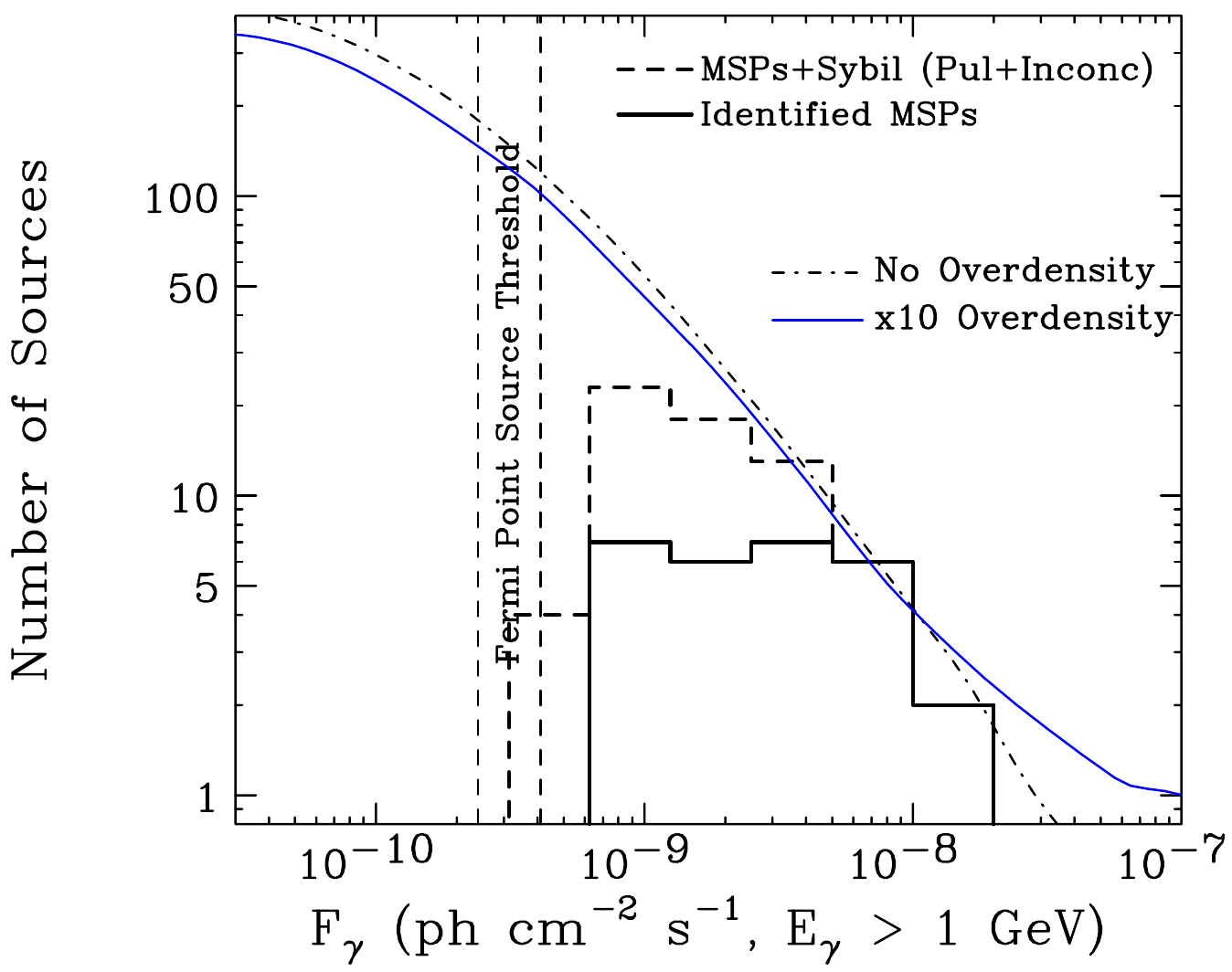}}
\mbox{\includegraphics[width=0.49\textwidth,clip]{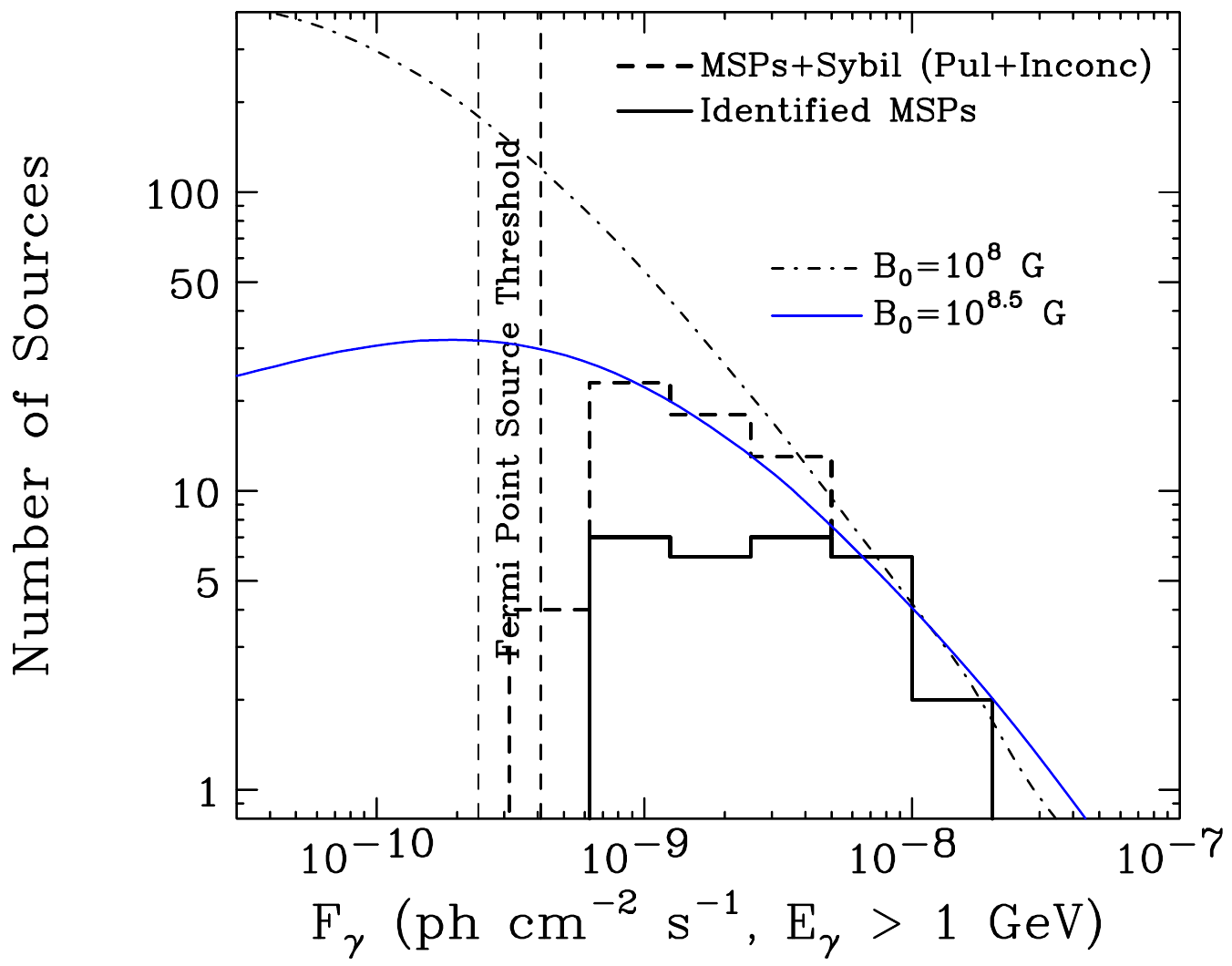}}
\caption{As in Fig.~\ref{histogram-dist}, but for a number of variations in the millisecond pulsar population model. See text for details.}
\label{histogram-vary}
\end{figure*}

To better accommodate the observed flux distribution of MSPs, we must consider population models with either 1) spatial distributions which are more weighted more toward nearby MSPs, or 2) luminosity functions that are more weighted more toward higher luminosity MSPs. In Fig.~\ref{histogram-vary}, we show how varying a number of our model's parameters can impact the flux distribution of MSPs. In the upper frames, we vary the parameters of our spatial distribution, $\langle|z|\rangle$ and $\sigma_r$ (see Eq.~\ref{spatial}). By reducing the vertical scale height of the MSPs distribution to $\langle|z|\rangle=0.3$ kpc (approximately the scale height of the Milky Way's thin disk), the model can provide a not unreasonable match to the observed distribution (although nearly all of {\it Sybil}'s non-AGN sources would have to be MSPs in this case). Values of $\langle|z|\rangle \gsim 1$ kpc appear to be incompatible with the observed flux distribution. In contrast, reasonable variations in $\sigma_r$ have relatively little impact on the predicted distribution. In the lower left frame, we consider the possibility of a local overdensity of MSPs (enhanced by a factor of 10 within 0.3 kpc of the Solar System). This, however, had little impact on the overall distribution, except for slightly increasing the predicted number of very bright sources. Lastly, in the lower right frame we focus on the MSP luminosity function by varying in the central value of the magnetic field distribution, $B_0$. We find that by increasing this quantity from $10^8$ to $10^{8.5}$ gauss or higher, we can much better accommodate the observed flux distribution. We also note that from the information shown in Fig.~\ref{PPdot}, values of $B_0 \sim 10^{8.3}-10^{8.5}$ G appear to best describe the observed population of MSPs.

\begin{figure*}
\mbox{\includegraphics[width=0.49\textwidth,clip]{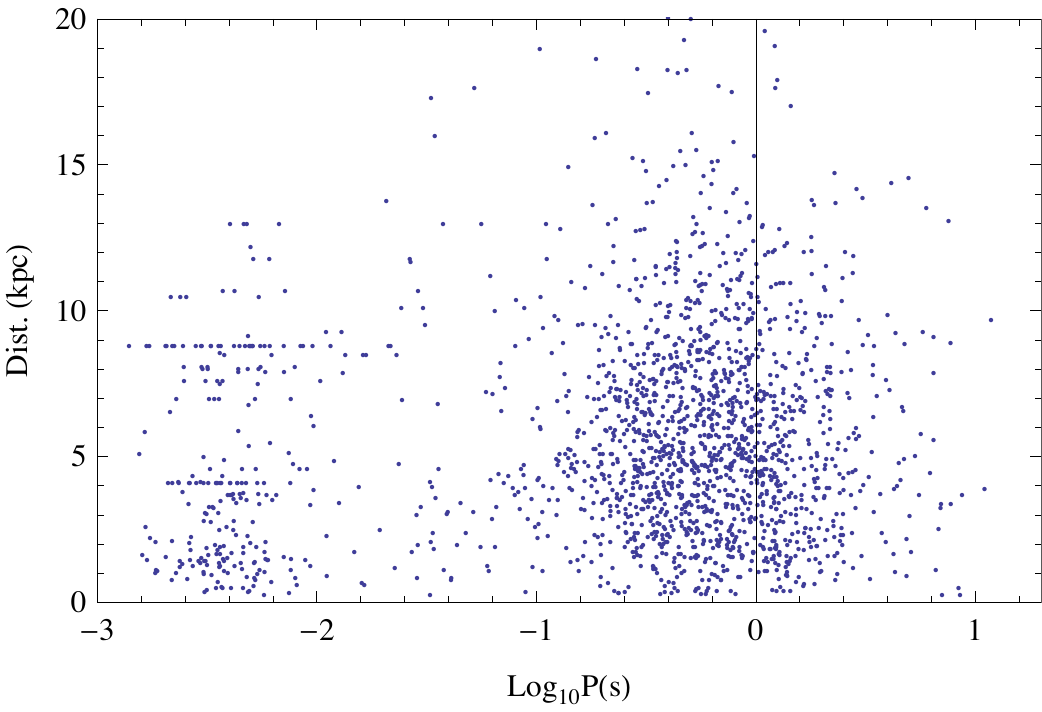}}
\mbox{\includegraphics[width=0.49\textwidth,clip]{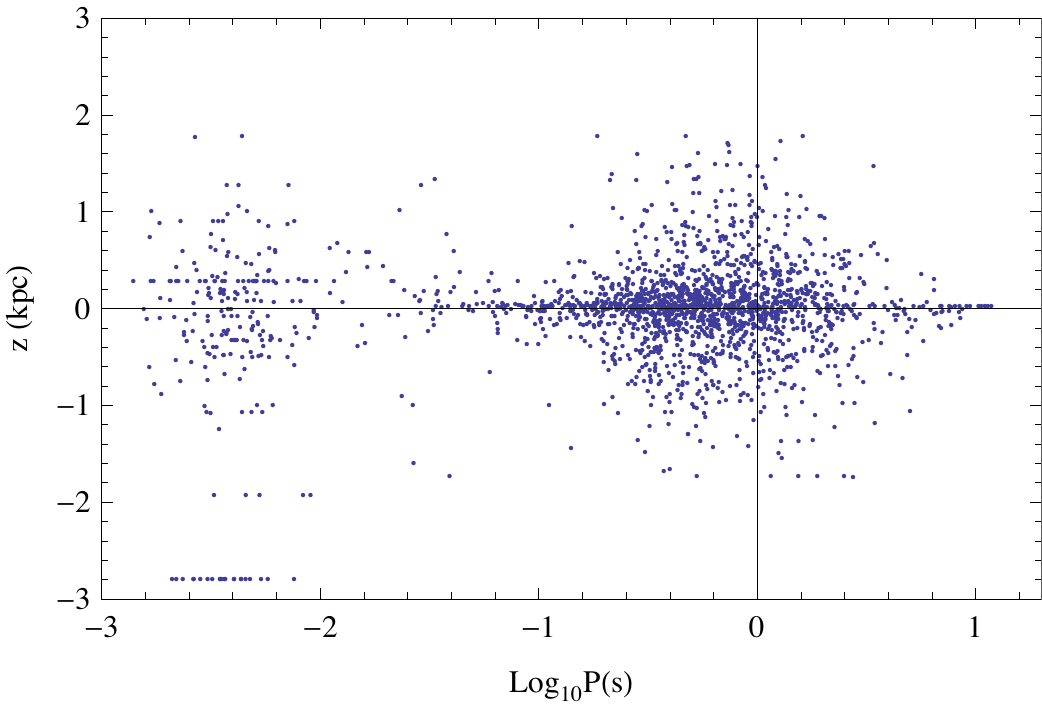}}
\caption{The distance from the Solar System (left) and distance from the Galactic Plane (right), as a function of period, for pulsars in the ATNF catalog~\cite{atnf}. The groups of points that form horizontal lines are millisecond pulsars in globular clusters.}
\label{maps2}
\end{figure*}

\begin{figure*}[!]
\mbox{\includegraphics[width=0.49\textwidth,clip]{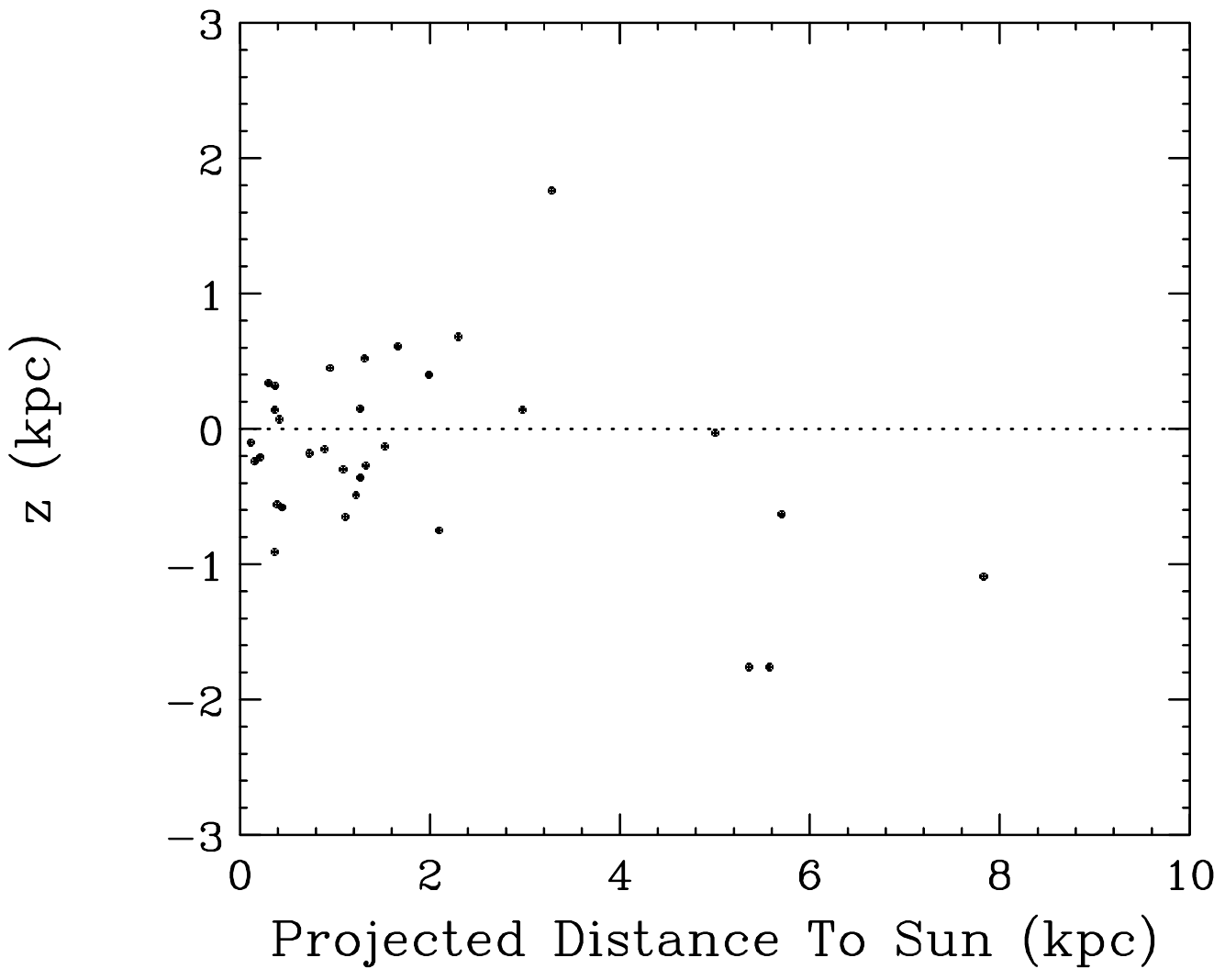}}
\mbox{\includegraphics[width=0.49\textwidth,clip]{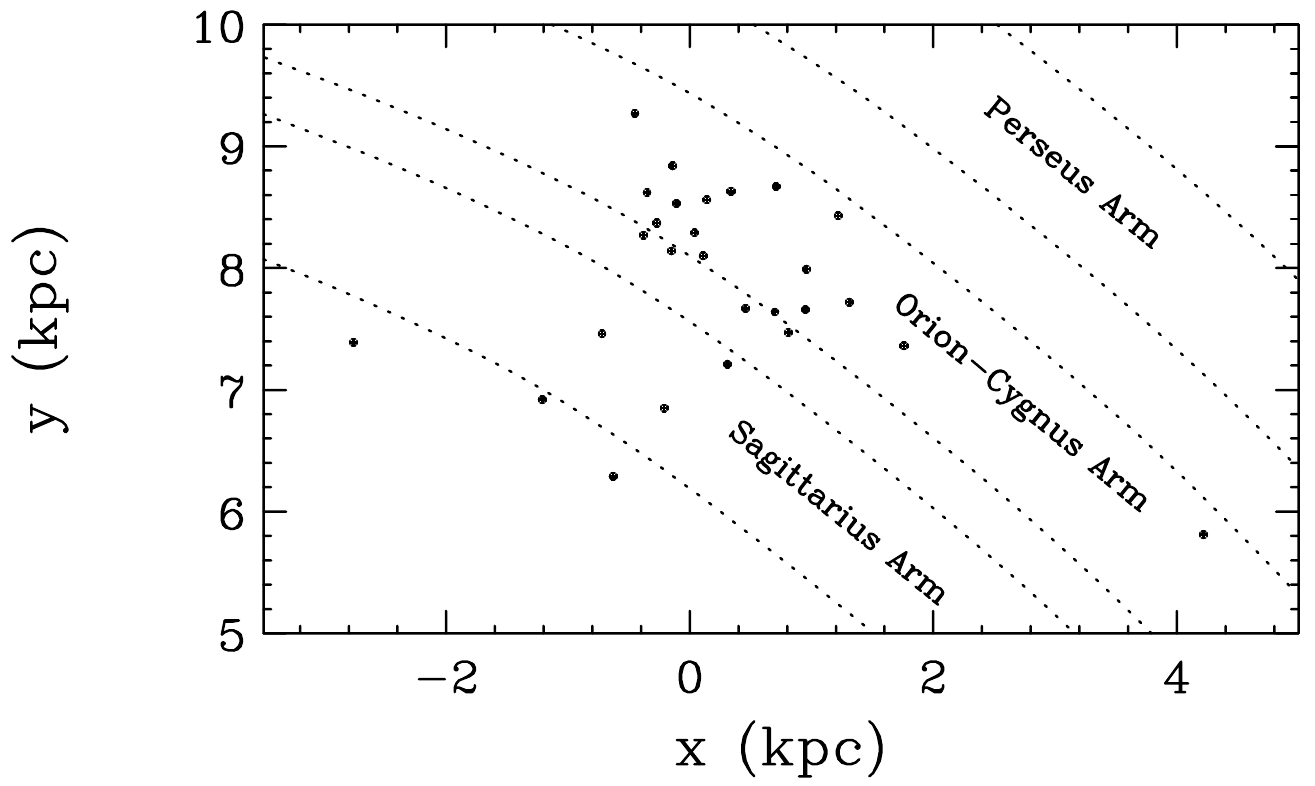}}
\caption{Left: The distance from the Galactic Plane and distance to the Solar System projected along the Galactic Plane for those millisecond pulsars observed by Fermi and with coordinates given in the ATNF catalog. Right: The locations in the Galactic Plane of those millisecond pulsars observed by Fermi and with coordinates given in the ATNF catalog, and with $|z|<1.5$ kpc. In this coordinate system, the Galactic Center is located at $(0,0)$, while the Solar System is at $(0,8.5 \, {\rm kpc})$. Shown for comparison are the approximate locations of the Orion-Cygnus, Sagittarius, and Perseus arms of the Milky Way.}
\label{maps}
\end{figure*}

\begin{figure*}
\mbox{\includegraphics[width=0.49\textwidth,clip]{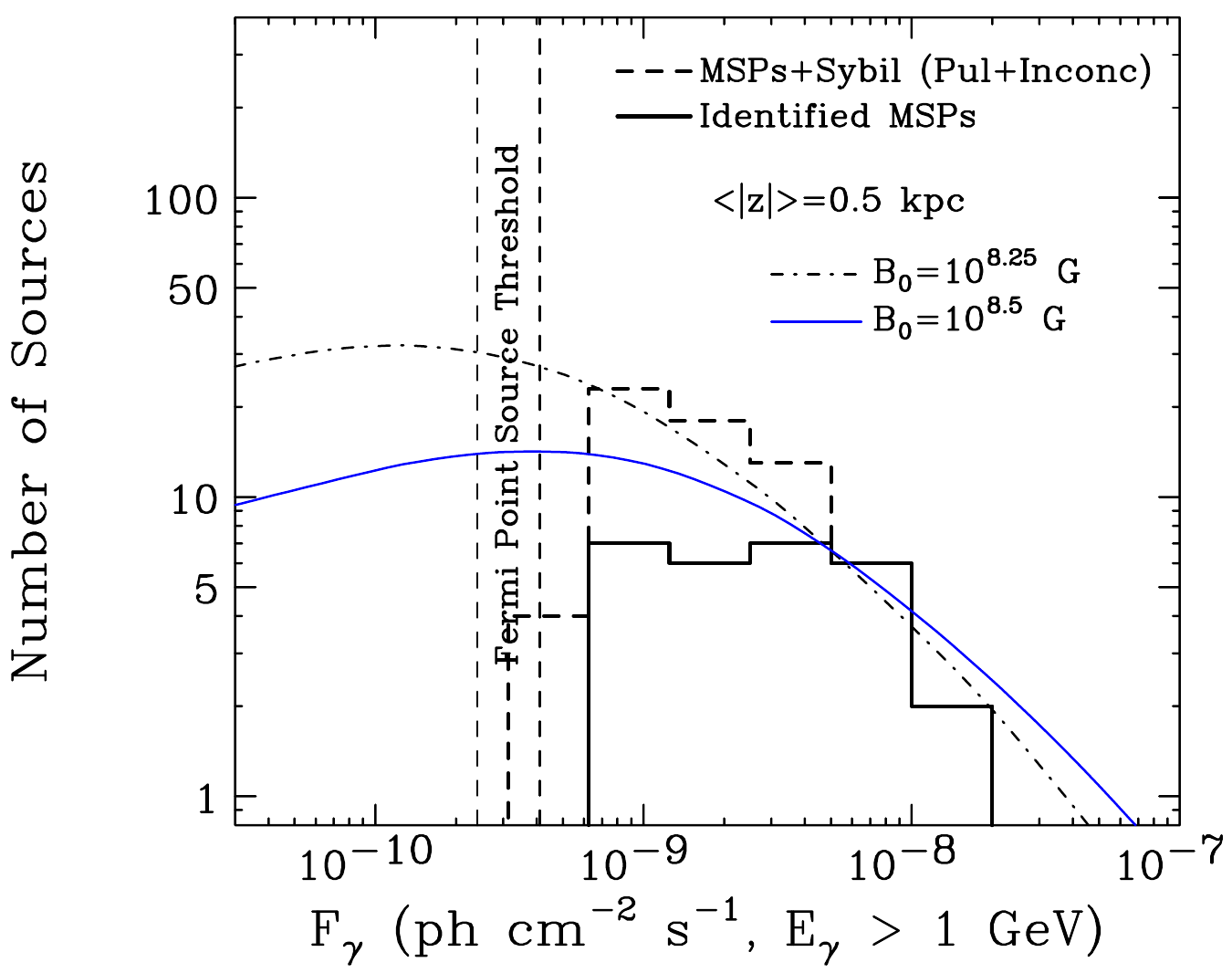}}
\mbox{\includegraphics[width=0.49\textwidth,clip]{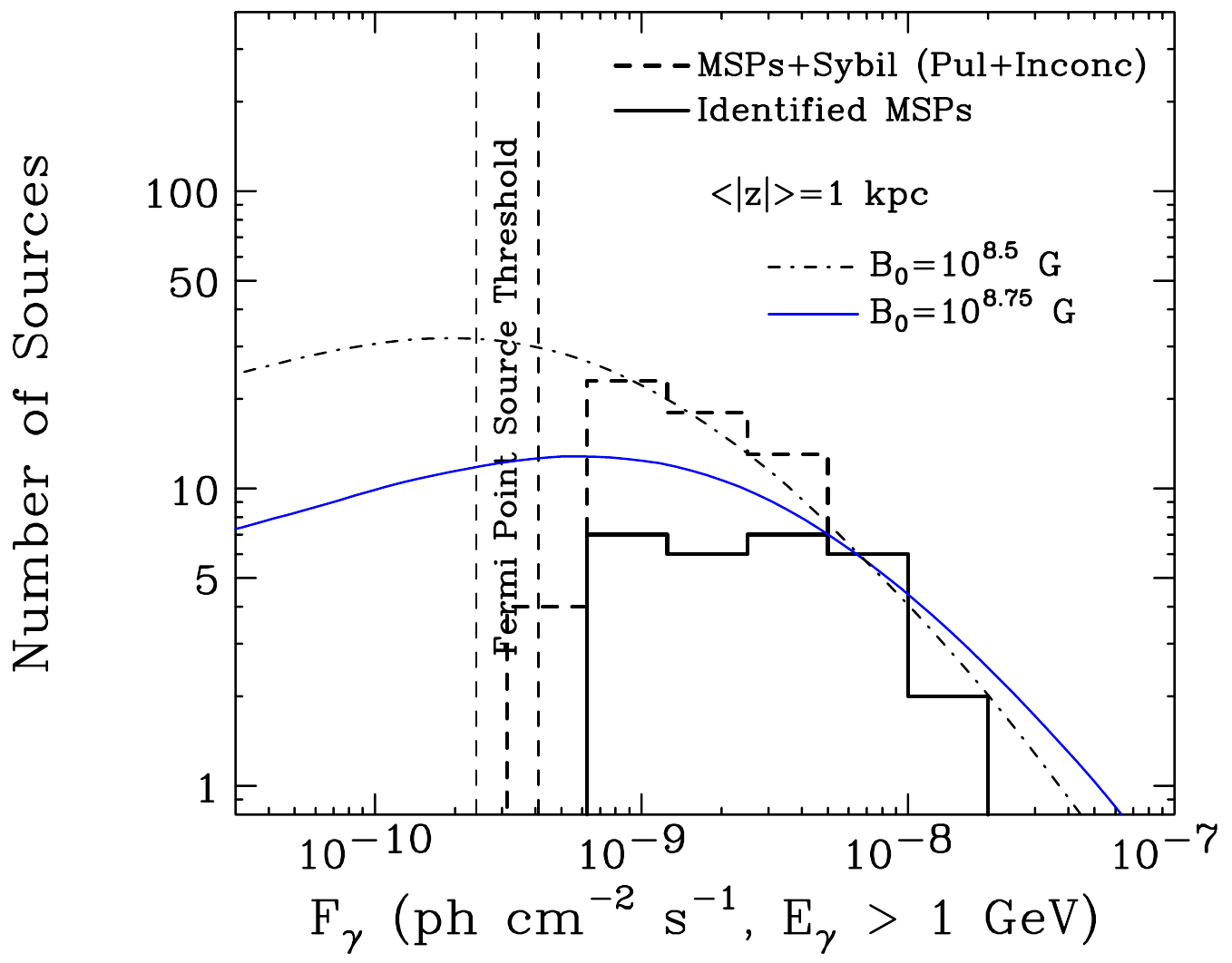}}\\
\mbox{\includegraphics[width=0.49\textwidth,clip]{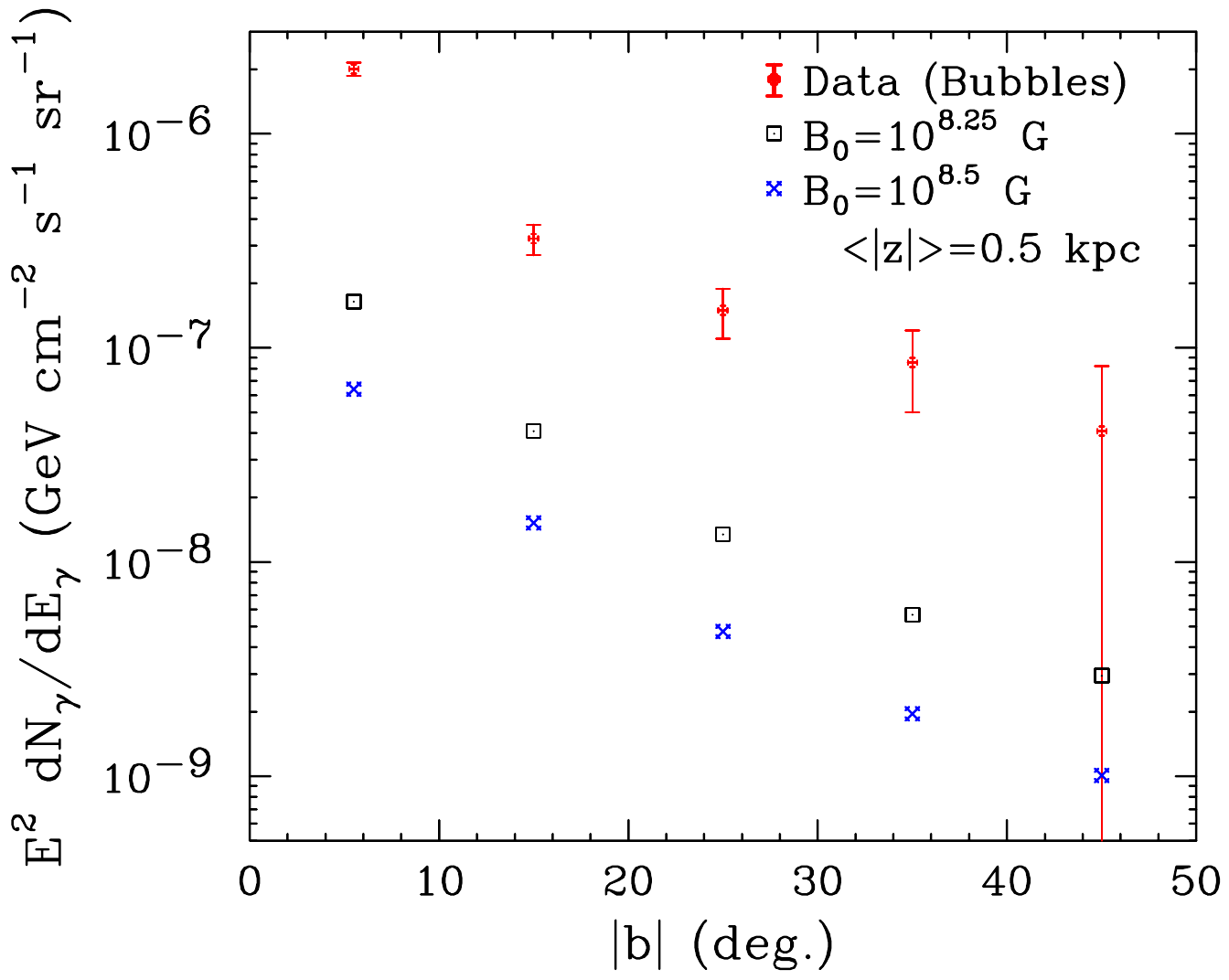}}
\mbox{\includegraphics[width=0.49\textwidth,clip]{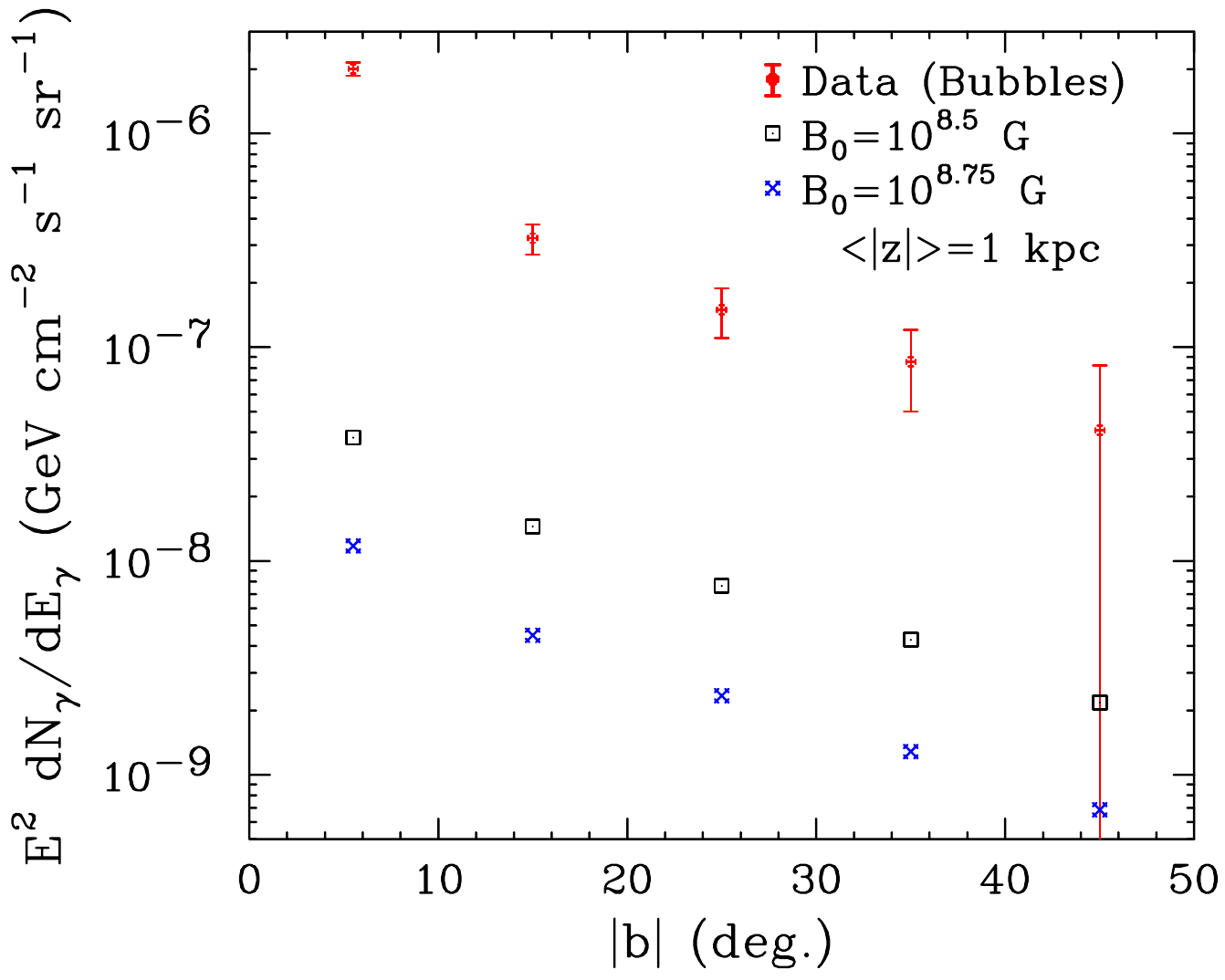}}
\caption{Top: As in Fig.~\ref{histogram-vary}, but for parameters which yield flux distributions which are in reasonable agreement with observations. Bottom: The observed gamma-ray flux (after subtracting inverse Compton emission) between 1.9 and 3.5 GeV from the regions associated with the Fermi Bubbles, in five latitude bands ($|b|=1^{\circ}-10^{\circ}$, $10^{\circ}-20^{\circ}$, $20^{\circ}-30^{\circ}$, $30^{\circ}-40^{\circ}$, and $40^{\circ}-50^{\circ}$), compared to the prediction from MSPs in the same four models used in the upper frames. In each case, only $\sim$5-10\% of the observed emission can be accounted for by millisecond pulsars. See text for details.}
\label{histogram-viable}
\end{figure*}

The spatial distribution of MSPs is not entirely unconstrained, however. In Figs.~\ref{maps2} and \ref{maps}, we plot some of the information we have pertaining to the spatial distribution of the MSPs observed at radio and gamma-ray wavelengths, respectively. Fig.~\ref{maps2} shows the distance to pulsars, and the distance of those pulsars from the Galactic Plane, as a function of period, for pulsars in the ATNF catalog~\cite{atnf}. The collections of points forming horizontal lines in these plots are groups of millisecond pulsars found in globular clusters. In Fig.~\ref{maps}, the spatial distribution of those MSPs observed by Fermi and with coordinates given in the ATNF catalog is shown. From these two figures, it is clear that the MSP distribution is not highly concentrated within a few hundred parsecs of the Galactic Plane, instead favoring a distribution at least as broad as $\langle|z|\rangle \simeq 0.5$ kpc.\footnote{While $\langle|z|\rangle \simeq 0.5$ kpc provides the best-fit to the observed distribution of observed MSPs, observational bias favoring nearby sources might lead us to slightly underestimate this quantity. We take 0.5 kpc to be an approximate lower limit for $\langle|z|\rangle$.} Thus reducing the value of $\langle|z|\rangle$ alone (as shown in the upper left frame of Fig.~\ref{histogram-vary}) does not seem to be a viable way to accommodate the observed flux distribution. Furthermore, the left frame of Fig.~\ref{maps2} and the lower frame of Fig.~\ref{maps} do not reveal any very large local overdensity of MSPs. 

From the results shown in Fig.~\ref{histogram-vary}, constrained by the observed spatial distributions of Figs.~\ref{maps2} and~\ref{maps}, we conclude that we are forced to consider MSP luminosity functions favoring somewhat higher values than are found in the FGL base model. In terms of the magnetic field parameter, this favors $B_0 \sim (2-6)\times 10^8$ G, although one should keep in mind that this parameter is somewhat degenerate with the period distribution, and with the fraction of rotational energy loss that goes into gamma ray production. We consider examples of what appear to be reasonably viable MSP population models in Fig.~\ref{histogram-viable}. In the upper frames, we show the flux distribution; for the lower choice of $B_0$ used in each frame (dot-dashed), we approximately saturate the observed source distribution. The distributions shown for slightly larger values of $B_0$ should be considered more realistic, many of the sources included in the dashed histogram are likely to be sources other than MSPs (in particular among {\it Sybil}'s inconclusive sources). In the lower frames of Fig.~\ref{histogram-viable}, we show the gamma-ray flux at 1.9-3.2 GeV (the approximate peak of the observed excess) observed by Fermi from various latitude ranges of the Inner Galaxy~\cite{tracy}, and compare this to the predicted flux in the same four MSP population models. Clearly these models cannot account for the observed emission, falling short in each case by a factor of $\sim$10-20.\footnote{In calculating the contribution to the diffuse gamma-ray emission as shown in the lower frames of Fig.~\ref{histogram-viable}, we have treated any MSP with a flux less than $4.1 \times 10^{-10}$ cm$^{-2}$ s$^{-1}$ above 1 GeV as unresolved and included its emission in the prediction for the diffuse flux. In light of the hard spectra of MSPs, this is a fairly conservative threshold, and we expect most MSPs with fluxes above $\sim2.4 \times 10^{-10}$ cm$^{-2}$ s$^{-1}$ to be resolved (see Fig.~6 of Ref.~\cite{catalog}). At high-latitudes, this provides a reasonable upper limit for the contribution to the diffuse flux. At lower-latitudes, however, some MSPs slightly brighter that our assumed threshold may go unresolved. If we increase our point source threshold by a factor of 2 (as is appropriate for sources at $|b|\simeq 10^{\circ}$~\cite{Abdo:2010ru}), we find that the low-latitude diffuse flux approximately doubles, still falling well short of that required to explain the observed emission.}

\begin{figure*}
\mbox{\includegraphics[width=0.49\textwidth,clip]{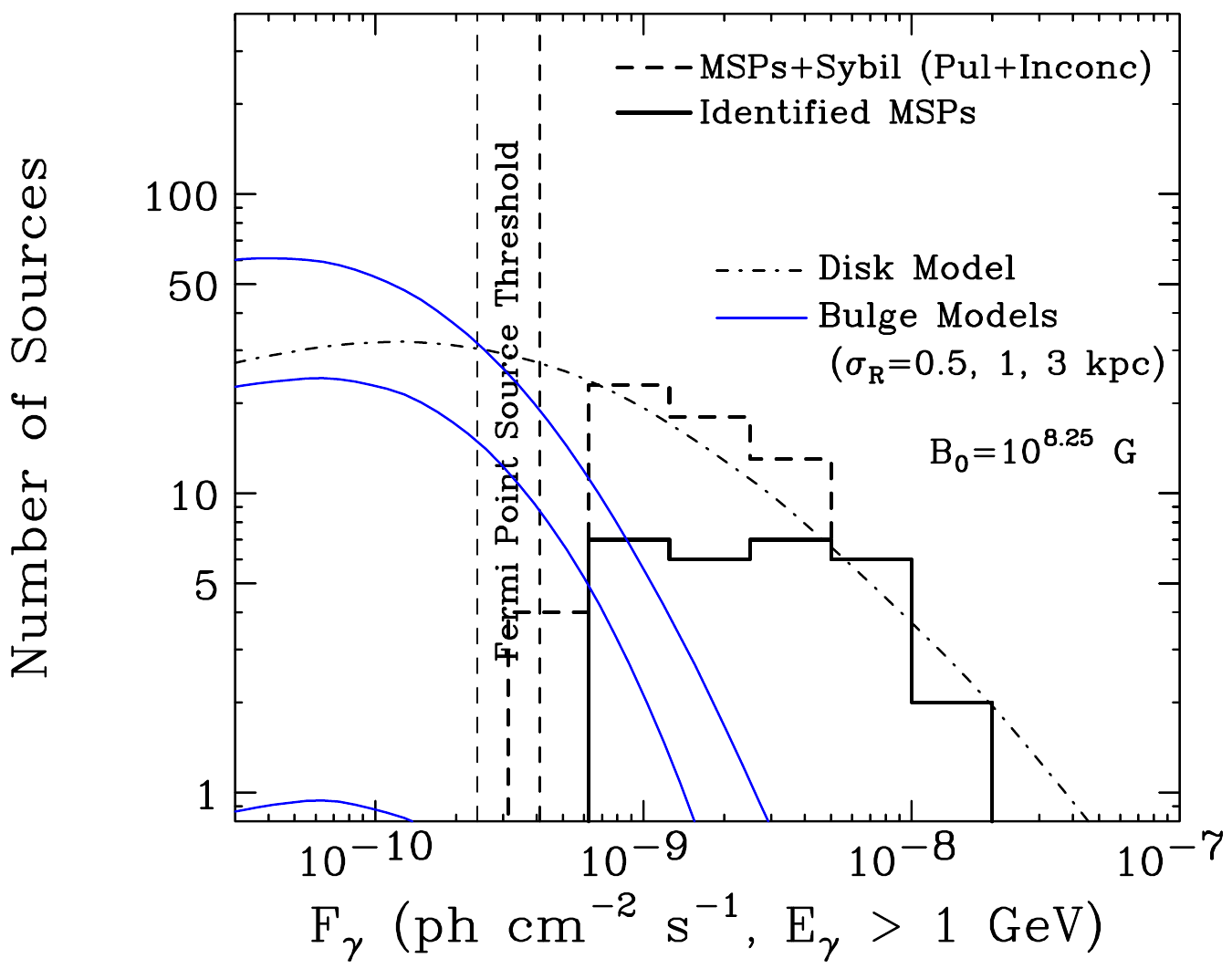}}
\mbox{\includegraphics[width=0.49\textwidth,clip]{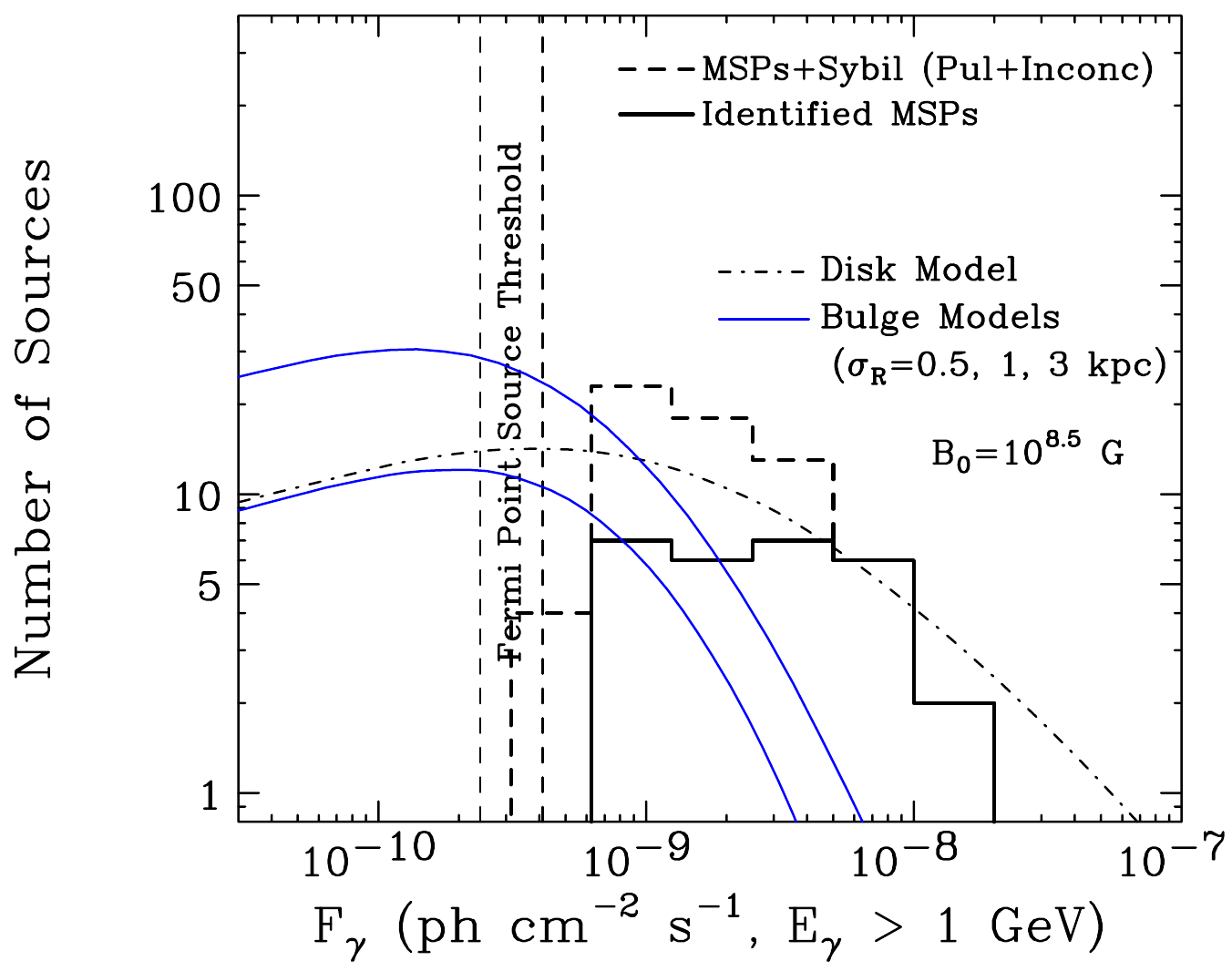}}
\caption{As in Figs.~\ref{histogram-dist},~\ref{histogram-vary} and~\ref{histogram-viable}, but now also showing the contribution from a population of millisecond pulsars associated with the Galactic Bulge. The three solid blue lines correspond to spatial distributions which are a spherical gaussian with $\sigma_R=0.5$, 1 and 3 kpc (from bottom-to-top, although the $\sigma_R=0.5$ contour falls below the range shown in the right frame). We have normalized the bulge contribution such that the number of millisecond pulsars per stellar mass is the same in the bulge as in disk. Here, we have also adopted a disk distribution with $\langle|z|\rangle=0.5$ kpc.}
\label{histogram-diskbulge}
\end{figure*}

\begin{figure*}
\mbox{\includegraphics[width=0.49\textwidth,clip]{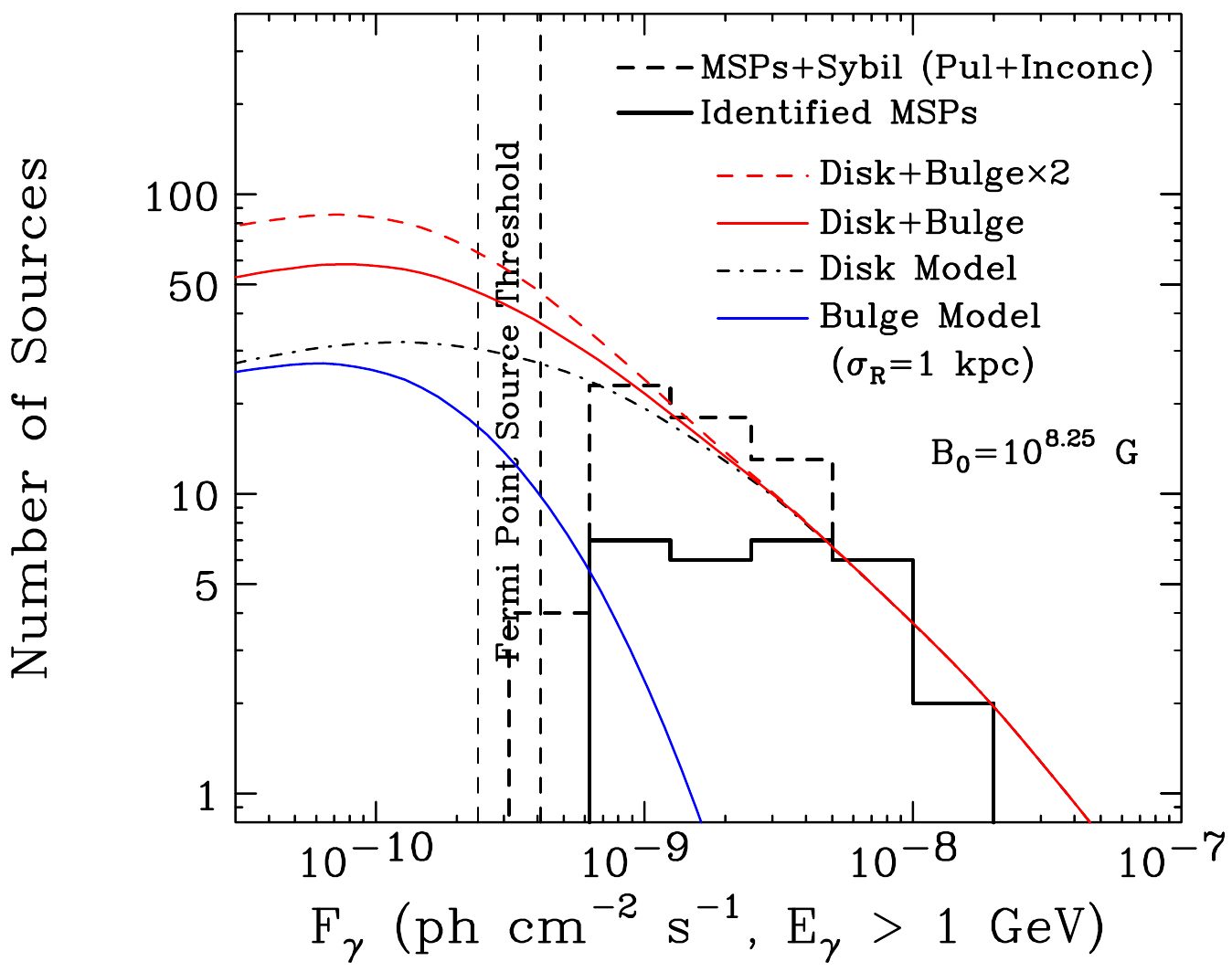}}
\mbox{\includegraphics[width=0.49\textwidth,clip]{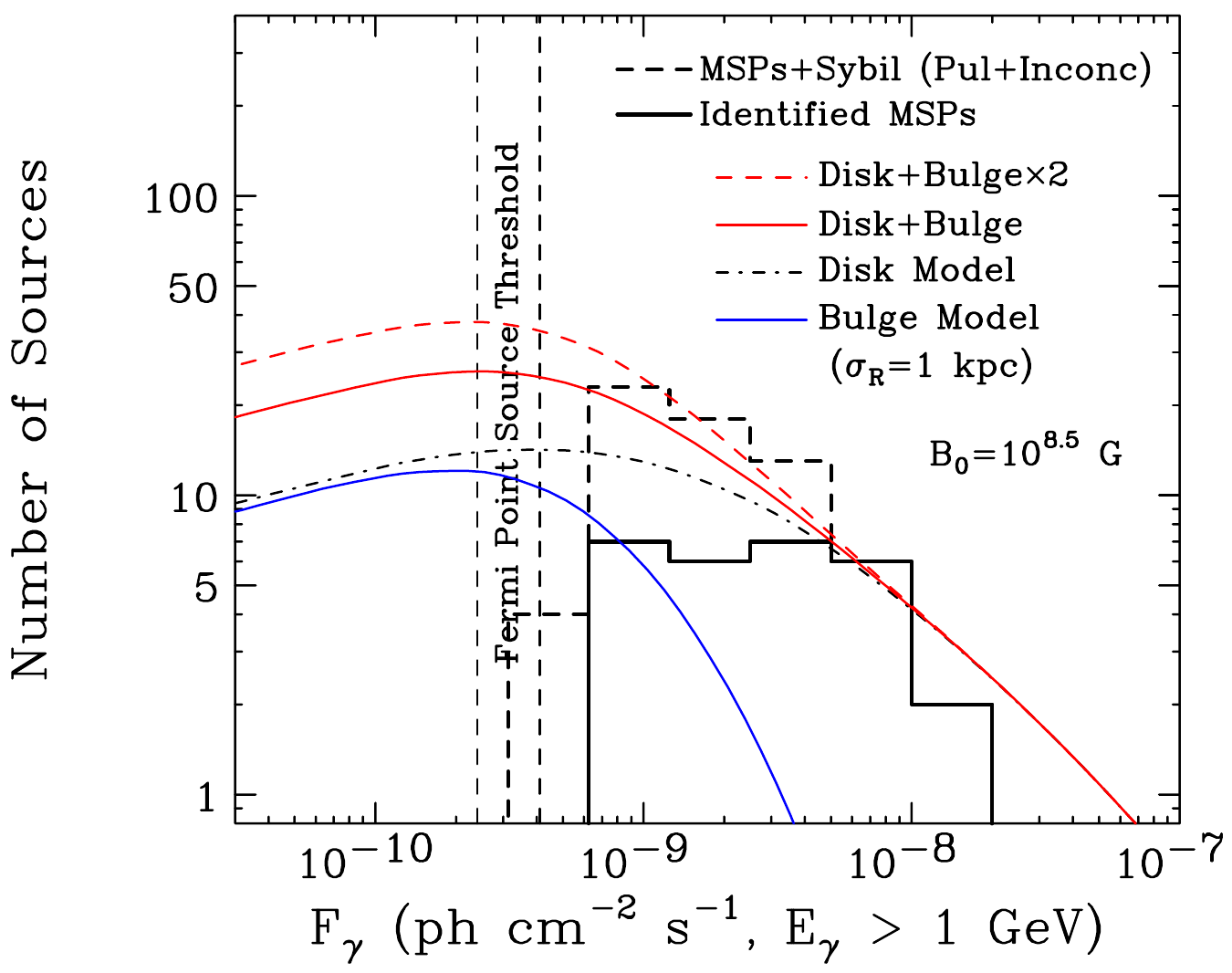}}\\
\mbox{\includegraphics[width=0.49\textwidth,clip]{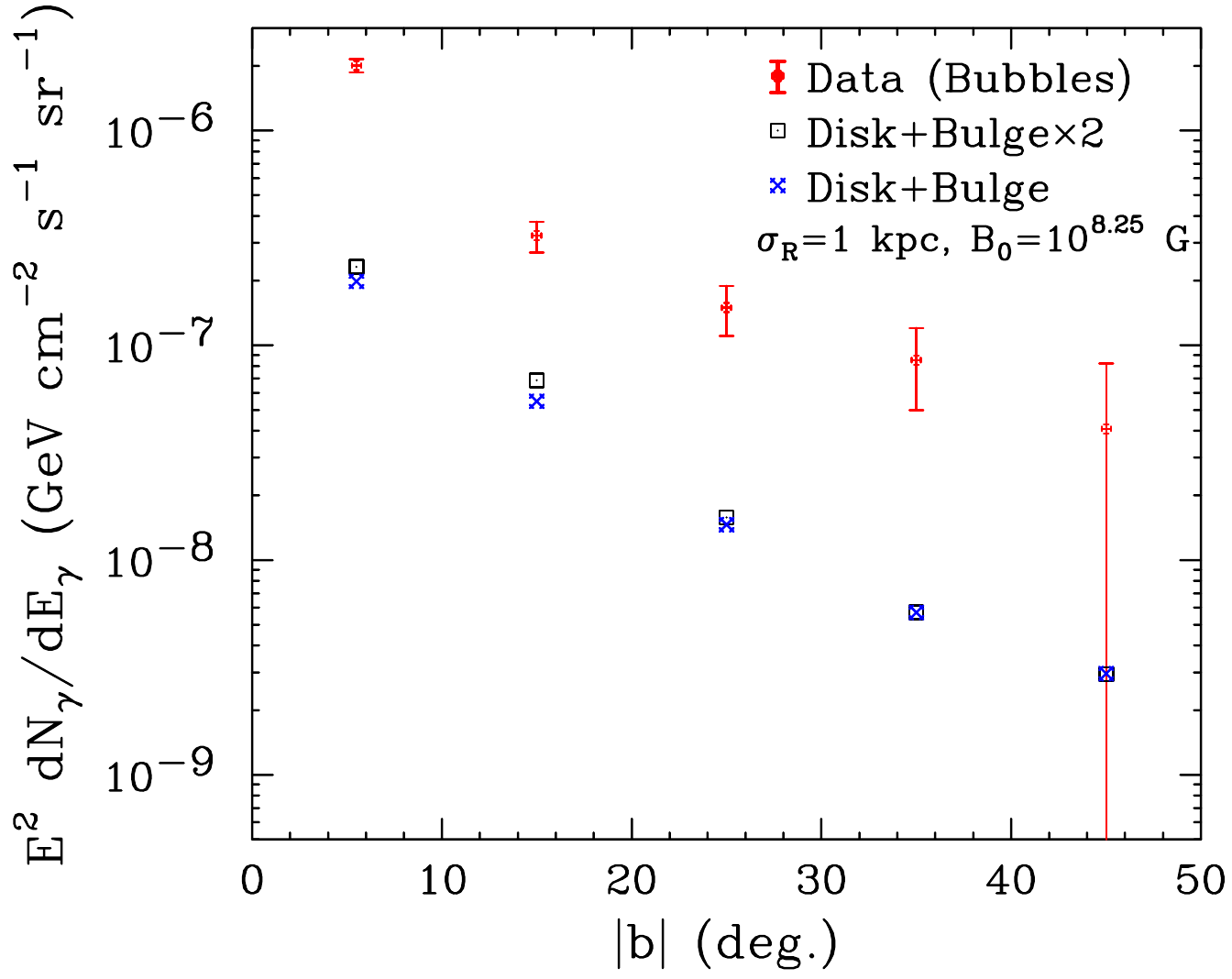}}
\mbox{\includegraphics[width=0.49\textwidth,clip]{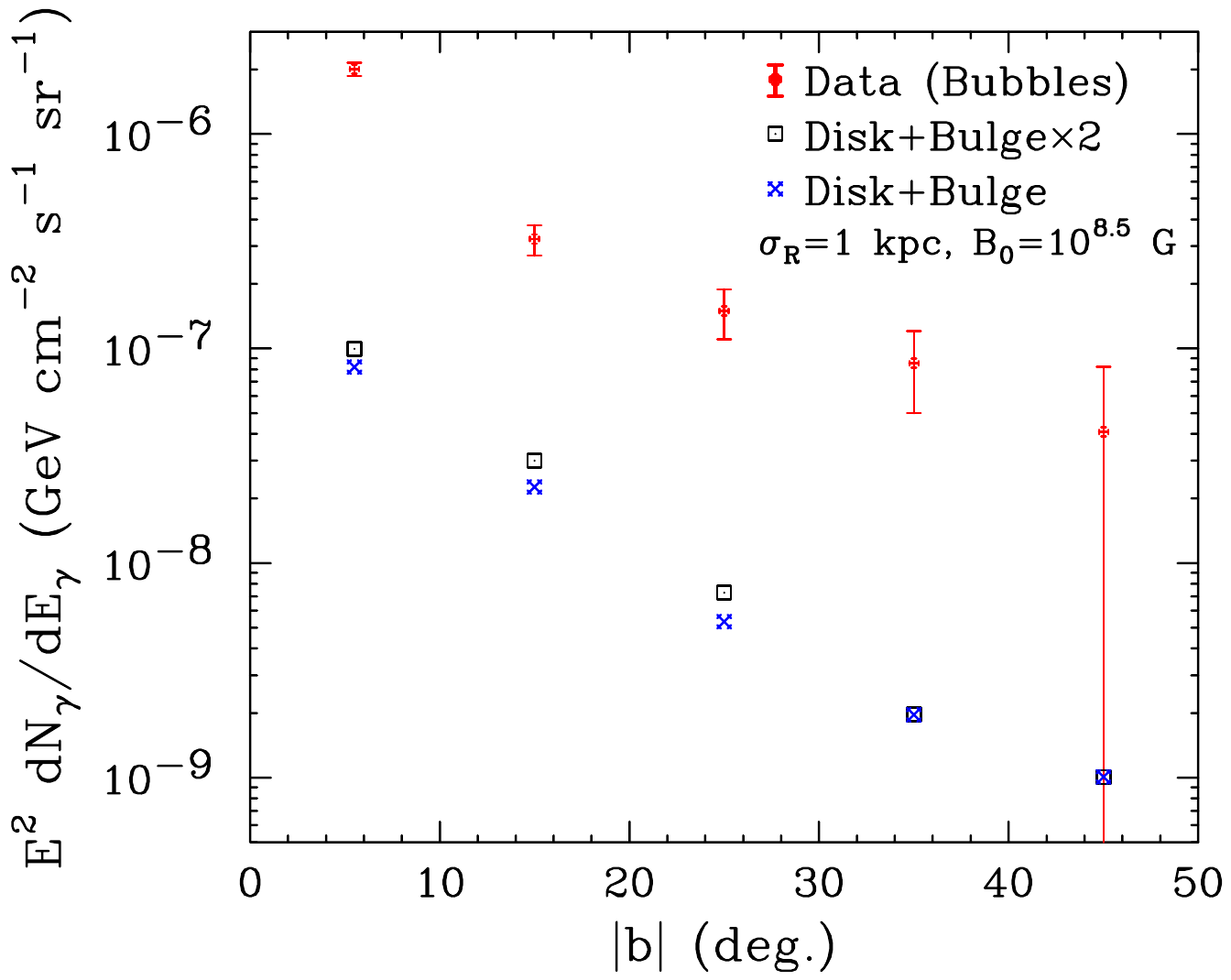}}
\caption{Top: As in Fig.~\ref{histogram-diskbulge}, but now also showing the bulge, disk, and bulge+disk contributions from millisecond pulsars. Here, we have adopted $\sigma_R=$1 kpc and $\langle|z|\rangle=0.5$ kpc. We have normalized the bulge contribution such that the number of millisecond pulsars per stellar mass is the same in the bulge as in disk (solid blue and solid red) and by a factor that is 2 times larger (dashed red). Bottom: As in the lower frames of Fig.~\ref{histogram-viable}, but for the sum of disk and bulge contributions. The total diffuse emission from millisecond pulsars is in each case found to be much less than that needed to account for the observed GeV excess.}
\label{histogram-diskbulgesum}
\end{figure*}


\subsection{Millisecond Pulsars Associated with the Galactic Bulge}

In the previous subsection, we showed that after taking into account the observed flux distribution and spatial distribution of MSPs, the population of MSPs associated with the Galactic Disk cannot produce more than $\sim$5-10\% of the Inner Galaxy's GeV excess. To increase the intensity of diffuse emission from unresolved MSPs in the Inner Galaxy without predicting far too many resolved sources, we must require an additional population of MSPs with a flux distribution that increases sharply below the point source sensitivity of Fermi. The most promising way to accomplish this is to add an additional component to our population model which explicitly takes into account those MSPs associated with the Galactic Bulge. In this subsection, we consider MSP models which include contributions from sources associated with both the disk and the bulge of the Milky Way.

We model the bulge population of MSPs with the same luminosity function as the disk component, and with a spatial distribution that is described by a spherically symmetric gaussian, $n(R)\propto \exp(-R^2/\sigma^2_R)$, where $R$ is the distance to the Galactic Center ($\sigma_R$ is not to be confused with the quantify $\sigma_r$, as appears in Eq.~\ref{spatial}). In Fig.~\ref{histogram-diskbulge} we show the flux distribution for disk and bulge components, using three values of $\sigma_R$, and for $B_0=10^{8.25}$ G (left), $B_0=10^{8.5}$ G (right), and $\langle|z|\rangle=0.5$ kpc. If we adopt a MSP distribution for the bulge that is similar to the distribution of bulge stars ($\sigma_R \simeq 0.5$ kpc), we get almost no contribution (the bottom blue curve barely appears in the left frame and falls below the range shown in the right frame, and thus does not appear). If we increase $\sigma_R$ to 1 kpc or more, we find a significantly larger contribution from the bulge, but also a non-negligible contribution to the number of individual sources that should be resolvable by Fermi. In particular, three of Fermi's observed MSPs exhibit gamma-ray luminosities of $2\times 10^{37}$ GeV/s or higher, each of which would be well above Fermi's point source threshold if located at a distance of $\sim$10 kpc from the Solar System (for $|b|>10^{\circ}$). The fact that a non-negligible fraction of bulge MSPs (those with high luminosities and outside of of the Galactic Plane) will be resolvable as individual point sources by Fermi will ultimately limit how much of the Inner Galaxy's GeV excess we can attribute to such a population.

The bulge contributions shown in Fig.~\ref{histogram-diskbulge} have been normalized assuming that the number of MSPs per stellar mass is the same in the bulge as in the disk. It is possible, however, that the relative number of MSPs in the bulge could be somewhat larger. As an extreme illustration of this possibility, we note that globular clusters are observed to contain $\sim$$10^2$ times more low mass X-ray binaries (the assumed progenitors of MSPs) per stellar mass than is found throughout the disk~\cite{gc1,gc2}. This is presumably the consequence of the very high stellar densities found in these systems (up to $\sim$$10^2$-$10^3$ stars per cubic parsec, compared to $\sim$0.4 in the local volume of the disk), which can be expected to significantly increase the probability that a given pulsar will obtain a companion and thus potentially evolve into a MSP. The average stellar density in the bulge is significantly higher than in the disk, but much lower than that found in the cores of globular clusters (only in the innermost tens of parsecs around the Galactic Center is the stellar density comparable to that found in globular clusters). As a result, we naively expect only a modest enhancement in the number of MSP per stellar mass found in the bulge relative to that in the disk (likely on the order of a few or less)~\cite{Wharton:2011dv,deneva}. This conclusion is further supported by the observed distribution of low mass X-ray binaries in the Galactic Bulge~\cite{Revnivtsev:2008fe}.

In Fig.~\ref{histogram-diskbulgesum} we show the distribution of sources (top) and flux of diffuse gamma-ray emission (as a function of latitude) from MSPs, including contributions from the disk and bulge. Here we have chosen a bulge distribution described by $\sigma_R=1$ kpc because significantly smaller values lead to a negligible contribution to the diffuse emission, while much larger values predict numbers of $\sim$$10^{-9}$ cm$^{-2}$ s$^{-1}$ sources that exceed those observed by Fermi. Furthermore, we find that our conclusions are not sensitive to the precise value of this parameter. We have also taken here a disk width of $\langle|z|\rangle=0.5$ kpc, which approximately maximizes the allowed contribution to the diffuse emission from MSPs in the bulge. In the lower frames of this figure, we show the gamma-ray flux at 1.9-3.2 GeV (the approximate peak of the observed excess) observed by Fermi from various latitude ranges of the Fermi bubbles~\cite{tracy}, and compare this to the predicted flux in these disk+bulge MSP population models. Again, we find that the predicted contribution from MSPs cannot account for the observed emission, falling short in each case by a factor of $\sim$10.




\section{Summary and Conclusions}
\label{summary}

Millisecond pulsars have been proposed as a possible source for the Inner Galaxy's GeV excess identified within the data of the Fermi Gamma-Ray Space Telescope. This hypothesis has been motivated by two main considerations. First, the spectrum of gamma-ray pulsars is observed to peak at $E_{\gamma}\sim$ 1-2 GeV, similar to that of the GeV excess. Second, it is plausible that the high stellar densities found in the Galactic Center could facilitate the production of a large number ($\sim$$10^3$) of millisecond pulsars, with a spatial distribution that is very highly concentrated within the innermost tens of parsecs of the Milky Way. It is far less clear, however, that such objects could account for the angular extent of this excess, which has recently been shown to extend out to at least $\sim$3 kpc from the Galactic Center~\cite{tracy}. In this paper, we address specifically this more extended signal, and the question of whether pulsars might account for this observed emission.

In considering the possibility that a large population of unresolved millisecond pulsars is the source of the Inner Galaxy's GeV excess, we have presented two independent arguments, each of which strongly disfavors this hypothesis:
\begin{itemize}
\item{In Sec.~\ref{spec}, we showed that the spectrum of the millisecond pulsars resolved by Fermi is not compatible with the observed spectral shape of the GeV excess. In particular, the combined spectrum of Fermi's 37 millisecond pulsars with spectral information contained in the 2FGL catalog is well fit by $dN_{\gamma}/dE_{\gamma} \propto E_{\gamma}^{-1.46} \exp(-E_{\gamma}/3.3 \,{\rm GeV})$, while the spectrum of the GeV excess is much harder at sub-GeV energies, {\it i.e.} $dN_{\gamma}/dE_{\gamma} \propto E_{\gamma}^{-0.5} \exp(-E_{\gamma}/2.75 \,{\rm GeV})$.}
\item{In Sec.~\ref{dist}, we considered models to describe the spatial distribution and luminosity function of millisecond pulsars in the Milky Way. After considering a wide range of parameters and distributions, we found no models that could accommodate the GeV excess without significantly overpredicting the observed number of bright, high-latitude ($|b|>10^{\circ}$) millisecond pulsars. Models that did not violate this constraint were capable of producing no more than $\sim$10\% of the GeV excess.}
\end{itemize}

In light of these results, we are forced to conclude that millisecond pulsars are not responsible for the GeV excess observed from the Inner Galaxy. Although one could imagine another (unknown) class of gamma-ray sources that could account for this signal, its members would be required to have a number of rather specific characteristics: 1) A very hard spectrum (much harder than pulsars), peaking at 2-3 GeV, 2) Low gamma-ray luminosities (to avoid being identified as individual point sources), consistently less than $\sim$$10^{37}$ GeV/s, and 3) A spatial distribution that is concentrated around the Inner Galaxy, with significantly more sources associated with the bulge than the disk (per stellar mass). No known class of gamma-ray sources exhibits these characteristics and, to the best of the authors' knowledge, no such class of sources has been proposed.

In excluding gamma-ray pulsars as the source of the Inner Galaxy's GeV excess, the results presented in this study provide further support for a dark matter interpretation of this signal. At this time, we know of no viable alternative to dark matter annihilations as the source of the excess GeV emission observed from the Galactic Center and from the inner kiloparsecs of the Milky Way. 


\acknowledgments{We would like to thank Manoj Kaplinghat, Kev Abazajian, and Albert Stebbins for helpful discussions. This work has been supported by the US Department of Energy. TRS is supported by the National Science Foundation under grants PHY-0907744 and AST-0807444. JSG acknowledges support from NASA through Einstein Postdoctoral Fellowship grant PF1-120089 awarded by the Chandra X-ray Center, which is operated by the Smithsonian Astrophysical Observatory for NASA under contract NAS8-03060.}

\bigskip
\bigskip
\bigskip
\bigskip
\bigskip
\bigskip

\begin{appendix}

\section{Kick Velocities and Millisecond Pulsars}
\label{kicks}

Stars massive enough to form a neutron star ($M \gsim 8 \, M_{\odot}$) become supergiants prior to collapse. If such a star is to leave its binary companion intact, the stars must be separated by a distance of at least an AU or so. With this picture in mind, we start by considering a binary system with masses $M_i$ and $M_{\rm comp}$ and in an orbit with a semi-major axis, $a$. The initial speed of the first star (which will become the pulsar), is given by $|\vec{V}_i|=\sqrt{G (M_i+M_{\rm comp})/a}$. Upon collapsing into a neutron star, the first star is given a kick velocity, $\vec{V}_{\rm kick}$. Adding these velocities, we arrive at $\vec{V}_f=\vec{V}_{\rm kick}+\vec{V}_{i}$. If $|\vec{V}_f| < |\vec{V}_i| \sqrt{2/\chi}$, the binary system will remain intact, potentially allowing for the future formation of a millisecond pulsar. The quantity $\chi=(M_i+M_{\rm comp})/(M_f+M_{\rm comp})$ takes into account the significant degree of mass loss that is expected to occur in explosion (typically $M_i\sim8-20\,M_{\odot}$ while $M_f\sim 1.4-3.2 \, M_{\odot}$).

As a numerical example, we consider a binary system with initial masses of 10 $M_{\odot}$ and 1.2 $M_{\odot}$, and in an orbit with a semi-major axis of 1 AU. Based on observed pulsar velocities, we take the distribution of initial kicks to be isotropic and maxwellian, with a mean velocity of 400 km/s. For these parameters, and for a final mass of the neutron star of $M_f=1.5\, M_{\odot}$, we find that only about 1\% of such systems will remain bound to each other (the other 99\% become ordinary pulsars, unless they later become bound to a new stellar companion). Of those pulsars which remain bound to their original companion, the mean kick velocity of the resulting binary system is only 27.6 km/s, which is sufficiently low to explain why so many observed MSPs are found within the relatively weak gravitational potentials of globular clusters. And while the details of this result depend on the input distributions of stellar masses, orbits, and initial kick velocities, the conclusion that MSPs should typically receive kick velocities that are smaller than those of ordinary pulsars by an order of magnitude or more is a generic prediction. In particular, for $a=1$ AU and $M_{\rm comp}=1.2 \, M_{\odot}$ we find average MSP kicks of 42 km/s for very massive neutron stars ($M_{i}=30\, M_{\odot}$, $M_f=3.2\,M_{\odot}$) and 26 km/s for minimally massive ($M_{i}=8\, M_{\odot}$, $M_f=1.4\,M_{\odot}$) neutron stars. Increasing $a$ or $M_{\rm comp}$ further reduces the average kick velocity. We do not consider much smaller values of $M_{\rm comp}$ than 1.2 $M_{\odot}$, as such stars would not have yet evolved into a giant phase, and thus would not have yet spun-up their neutron star companion into a MSP.

We also note that it has been suggested that the distribution of pulsar kick velocities may be bi-modal. Specifically, electron-capture supernovae are thought to often result in pulsars with relatively weak kick velocities~\cite{bimodal}. This could significantly enhance the fraction of pulsars which retain a binary companion and thus potentially evolve into a MSP.

\end{appendix}

\end{document}